\documentclass[pre,preprint,superscriptaddress]{revtex4-1}

\usepackage{amsmath}  
\usepackage{amsfonts} 
\usepackage{graphicx} 
\usepackage{enumitem} 
\usepackage{xcolor} 
\usepackage[utf8]{inputenc} 
\usepackage{tikz} 
\usepackage{csquotes}

\usepackage[subrefformat=parens,labelformat=parens]{subfig}

\begin{document}


\title{Markov chain approach to anomalous diffusion on Newman-Watts networks}

\author{Alfonso Allen-Perkins}
\email{alfonso.allen@hotmail.com}
\affiliation{Instituto de F\'isica, Universidade Federal da Bahia, 40210-210 Salvador, Brazil.}
\affiliation{Complex System Group, Universidad Polit\'ecnica de Madrid, 28040-Madrid, Spain.}%

\author{Alfredo Blanco Serrano}
\email{alfredoblancoserrano@gmail.com}
\affiliation{Instituto de F\'isica, Universidade Federal da Bahia, 40210-210 Salvador, Brazil.}%

\author{Thiago Albuquerque de Assis}
\email{thiagoaa@ufba.br}
\affiliation{Instituto de F\'isica, Universidade Federal da Bahia, 40210-210 Salvador, Brazil.}%
\affiliation{Complex System Group, Universidad Polit\'ecnica de Madrid, 28040-Madrid, Spain.}

\author{Juan Manuel Pastor}
\email{juanmanuel.pastor@upm.es}
\affiliation{Complex System Group, Universidad Polit\'ecnica de Madrid, 28040-Madrid, Spain.}

\author{Roberto F. S. Andrade}
\email{randrade@ufba.br}
\affiliation{Instituto de F\'isica, Universidade Federal da Bahia, 40210-210 Salvador, Brazil.}%

\date{\today}

\begin{abstract}

A Markov chain (MC) formalism is used to investigate the mean-square displacement (MSD) of a random walker on Newman-Watts (NW) networks. It leads to a precise analysis of the conditions for the emergence of anomalous sub- or super-diffusive regimes in such random media. Whereas results provided by most numerical approaches used so far base their results on the computation of a large number of independent runs over many equivalent substrates, the MC framework is applied only once to each equivalent sample. Starting from the simple cycle graph with $2k$ nearest neighbor connections, for which exact MSD expressions within the MC formalism can be derived, the randomness and complexity of the substrate is easily controlled by the number $x$ of added links. Results for different values of $k$, $x$, and the number $N$ of nodes make it possible to distinguish actual anomalous regimes from transient behavior and finite size effects. Albeit the high computing cost restricts the size of our networks to $N\leq1500$ nodes, our very precise results justify a new and more comprehensive scaling ansatz for walker dynamics, from which the behavior for very large networks can be derived.

\vspace{0.25cm}


\end{abstract}

\maketitle

\section{Introduction}


Diffusion dynamics in networks has becomes a most investigated issue, impacting several areas of knowledge, from epidemics \cite{satorras15} to social behavior \cite{centola10,valente17}, from economics \cite{kenett15} to world-wide goods transportation \cite{caldareli12,rozenblat08}. For several classes of complex networks, diffusive processes present striking differences to similar phenomena on the regular Euclidian lattices. For most of them, such changes can be explained by the fact that the average shortest-path between their nodes increases at most logarithmically with the number of nodes. This is the case of the small world (SW) networks, characterized by a small average shortest-path as well as by a large clustering coefficient. The first mathematical models to generate networks with these effects were those proposed by Watts and Strogatz (WS) \cite{watts98} and Newman and Watts (NW) \cite{Newman99}, whose topological properties were extensively investigated in early studies \cite{Newman00,kulkarni00,almaas02}. Later on, other models were also discussed in the literature \cite{dorogovtsev00,porter12}.

The WS and NW networks have also been intensively used to investigate several aspects of random walks on SW structures, e.g.: spectral properties of the Laplacian operator \cite{monasson99}, computer simulation based analyzes of mean number of visited sites and the return probabilities \cite{jespersen00,almaas03,lahtinen01},  analytic results for the average access time \cite{pandit01}, target problems \cite{jasch01}, and mean traversal time \cite{parris05}. Joining together analytical results to numerically obtained insights and evidences, it became possible to obtain certain scaling relations involving space and time. Nevertheless, it is clear that such relations are strongly dependent on the network structure as well as on the different walking strategies \cite{yang05}. A recent review summarizes most of research results on this issue \cite{Masuda17}.

Despite these advances, some crucial aspects of the behavior of the system still remain not well settled as, for instance, the time dependence of the mean-square displacement (MSD) of a random walker. Previous studies of MSD on NW networks identify the same regular (Gaussian) behavior MSD $\sim t$  during the initial steps of the walk \cite{almaas02,almaas03}. On the other hand, indications that super-diffusion may emerge when some topological features of SW networks are changed have been reported \cite{Huang06}. To the best of our knowledge, no single study provides a clear discussion for the coexistence of different regimes and the transition from one regime to the other.

In this work we resume the analysis of scaling properties of MSD on NW networks using an exact Markov chain (MC) formalism. Its results contrast with those provided by a large number of studies, which are obtained by averaging over a very large number of numerical simulations for different network realizations and initial conditions. By averaging only over different samples, the MC approach avoids intrinsic large fluctuations of the individual runs. Because of this aspect, it becomes possible to  obtain a precise identification of combined sub- and super-diffusive behaviors, as well as transient and finite size effects, among the rich variety of diffusive patterns.

Networks of the NW class depend on the parameters $N$, $k$, and $x$. They correspond, respectively, to the number of nodes in the cycle graph ($N-$ring), to half the number of nearest neighbors originally connected to each node, and the number of randomly added links (shortcuts). Therefore, starting from a regular cycle-graph ($k = 1,\,x = 0$), we study in detail the dependence of MSD on $k$ and $x$ as well as on $N$, advancing beyond the sparse NW network regime ($p\equiv x/Nk\ll 1$). Effects of the topological structures are already important when we depart from limiting case, so that sub- or super-diffusive behavior appear already when $k>1$ and or $x >0$. The crossover times between diffusive regimes and saturation depend on $x$ and $N$.  As MC approach relies on matrix operations, it becomes prohibitive to consider extreme large networks, as considered in usual numerical simulations. Nevertheless, working with medium-size networks $N\lesssim 1500$, it is possible to suggest a highly precise scaling ansatz for MSD, which extends itself to $ 0\leq p \lesssim 0.1$.

The paper is organized as follows. In Sec.~\ref{MSD_markov}, we introduce the Markov chain framework. Section \ref{arrangements} defines variants of the cycle graph that result in NW-networks by the sequential inclusion of shortcuts. Section \ref{results} presents our MC results for MSD on the networks discussed in Sec.~\ref{arrangements}, and draws a comparison with scaling properties found by numerical simulations for walks in large size NW-networks. Finally, our conclusions are summarized in Sec.~\ref{conclusiones}.


\section{Markov Chain formalism for evaluation of MSD in networks}
\label{MSD_markov}

According to Refs.~\cite{Newman99, Newman_S_W_01}, the NW-networks are simple, undirected graphs without self-loops. In such systems, the usual \textit{discrete time random walk} (or Polya walk) is a random sequence of vertices generated as follows: given a starting vertex $i$, denoted as \enquote{origin} of the walk, at each discrete time step $t$, the walker jumps to one nearest
neighbor of its current node \cite{Masuda17,aldous02,lovasz93}.

The MSD, defined by $\left \langle r^2(t) \right \rangle$, is a measure of the ensemble average distance between the position of a walker at a time $t$, $x(t)$, and a reference position, $x_0$. Assuming that $\left \langle r^2(t) \right \rangle$ has a power law dependence with respect to time, we have:

\begin{equation}
\mathrm{MSD}\equiv \left \langle r^2(t) \right \rangle = \left \langle \left ( x(t)-x_0 \right )^2 \right \rangle \sim  t^\gamma,
\end{equation}

\noindent where the value of the parameter $\gamma$ classifies the type of diffusion into normal diffusion ($\gamma=1$), subdiffusion  ($\gamma<1$), or superdiffusion ($\gamma>1$). Although MSD is one of the used measures to analyze general stochastic data \cite{almaas03,gallos04}, in order to better characterize diffusion, additional measures are also required, e.g., first passage observables \cite{Masuda17}. For the type of results we discuss here, $\left \langle r^2(t) \right \rangle$ is essential to provide a clear cut way to characterize the time dependence.

Following Refs.~\cite{Masuda17,estrada17b,zhang13}, we start the study of $\left \langle r^2(t) \right \rangle$ on simple networks, by considering an analytical expression for the probability of finding a random walker at a given node at time $t$, when the random walker is initially located at node $i$.

Let $G=(V,E)$ be a simple, undirected graph or network without self-loops, whose adjacency matrix is denoted by $\mathbf{A}$: $\mathbf{A} \left (i,j \right ) = 1$ if vertices $i$ and $j$ are connected, and $\mathbf{A} \left (i,j \right ) = 0$ otherwise. Let us define the strength (or degree) of a given node $i$ as:

\begin{equation}
s \left ( i \right )=\left ( \mathbf{A} \vec{1} \right )_i=\sum_{j=1}^{N}\mathbf{A} \left (i,j \right )
\end{equation}

\noindent where $\vec{x}$ is an all-x vector. Consequently, the probability that a particle staying at node $i$ moves to the node $j$ is given by:

\begin{equation}
P \left ( i,j \right )= \frac{\mathbf{A} \left (i,j \right )}{s \left ( i \right )}.
\label{prob_elem_matr}
\end{equation}

Let us denote by $\mathbf{\mathcal{S}}$ the diagonal matrix with elements $\mathbf{\mathcal{S}} \left ( i,i \right )=s \left ( i \right )$ and let us define the \textit{transition matrix} for the random walk as $\mathbf{\mathcal{P}}= \mathbf{\mathcal{S}}^{-1}\mathbf{A}$, whose elements are given by Eq.~\ref{prob_elem_matr}. According to this definition, ${\mathbf{\mathcal{P}}}$ is a stochastic matrix and the random-walk has the Markov property: the conditional probability distribution of future positions of the walker depends only upon the present position, not on the sequence of nodes that preceded it. Thus, the vector $\vec{p}_t$ describing the probability of finding a random walker at a given node of the graph at time $t$ evolves with time according to:

\begin{equation}
\vec{p}_{t+1}= { \mathbf{\mathcal{P}}} ^T \vec{p}_{t},
\end{equation}

\noindent where $\mathbf{X}^T$ stands for the transpose of matrix $\mathbf{X}$.

The vector $\vec{p}_{t}$ depends on the initial position $\vec{p}_0$ of the walker. We denote by $\vec{p}_{t,i}$ the vector containing the probability of finding the walker at a given node of the graph at time $t$, when the walk starts at node $i$. Therefore, it is possible to write the following expression:

\begin{equation}
\vec{p}_{t,i}=\mathbf{\mathcal{P}}_{t-1}^T \cdots \mathbf{\mathcal{P}}_{0}^T \vec{p}_{0,i}= \left (\mathbf{\mathcal{P}}^T \right )^t \vec{p}_{0,i},
\label{time_evolution_prob}
\end{equation}

\noindent where $\left ( \vec{p}_{0,i} \right )_j=1$ if $i=j$, and $0$ otherwise.

With the help of Eq.~(\ref{time_evolution_prob}) it is possible to quantify the mean distance $r$ covered by a typical walker. Thus, given $G=(V,E)$ and an initial condition $\vec{p}_{0,i}$, we calculate $\left \langle r^2(t) \right \rangle$ of the random walker to the origin (i.e., the node $i$), at each time step, $r^2(t,i)$, as follows:

\begin{equation}
r^2(t,i)=\sum_{j=1}^{N} \left ( d_{i,j} \right )^2 \left ( \vec{p}_{t,i} \right )_j,
\label{MSD_una_condicion_new}
\end{equation}

\noindent where $d_{ij}$ is the length of the shortest path distance between $i$ and $j$, that is, the smallest number of edges connecting the nodes $i$ and $j$. To obtain numerical estimates for $\left \langle r^2(t) \right \rangle$, we average over all the different initial positions of the walker:

\begin{equation}
\left \langle r^2(t) \right \rangle = \frac{1}{N}\sum_{i=1}^N r^2(t,i)= \frac{1}{N}\sum_{i=1}^N \sum_{j=1}^{N}\left ( d_{i,j} \right )^2 \left ( \vec{p}_{t,i} \right )_j.
\label{eq_MSD_multpl_new}
\end{equation}

\noindent According to Eqs.~\ref{time_evolution_prob} and \ref{eq_MSD_multpl_new}, in the case of a simple undirected networks without self-loops, $\left \langle r^2(t) \right \rangle$ only depends on the discrete time-step $t$, and the topology of $G$, given by $\mathbf{A}$.

Finally, note that the MC methodology provides numerically exact results for $\left \langle r^2(t) \right \rangle$. However, this method involves the use of matrices (see Eq.~\ref{eq_MSD_multpl_new}) and, in the case of large networks, these calculations are CPU and time demanding. For that reason,the results we discuss in this work are limited to $N\leq1500$, which is however large enough to reach convergence to the $N\rightarrow \infty$ for the majority of our results.


\section{From cycle graphs to NW-networks}
\label{arrangements}

Let us consider a regular one-dimensional ring with $N$ nodes that have nearest- and next-nearest-neighbor connections out to some constant range $k$ (i.e., each node is initially connected to its $2k$ nearest neighbors), and periodic boundary conditions. Hereafter, we will denote such a systems as $(N,k)-$\textit{cycle graphs}, $C_{N,k}$. Note that, in graph theory, the term \textit{cycle graph} usually refers to the case $C_{N,1}\equiv C_{N}$.

\begin{figure}[h!]
\centering
\subfloat[]{
\includegraphics[width=0.5\linewidth]{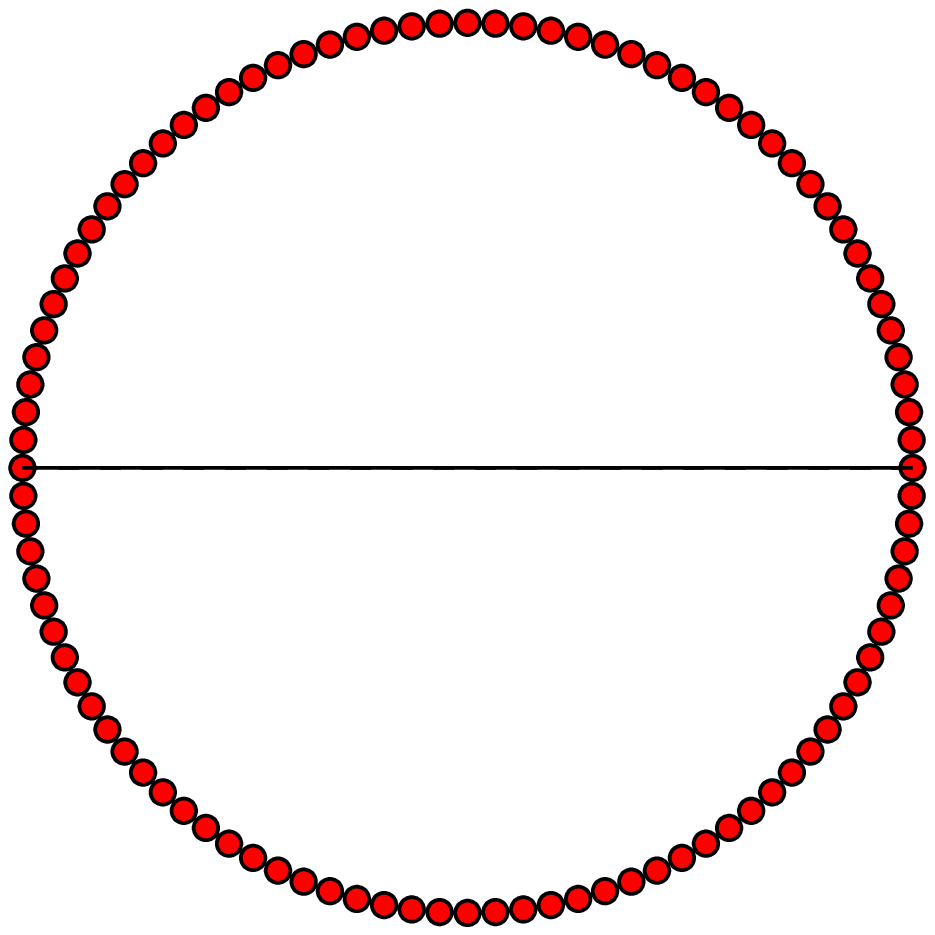}
\label{offset0}
}
\subfloat[]{
\includegraphics[width=0.5\linewidth]{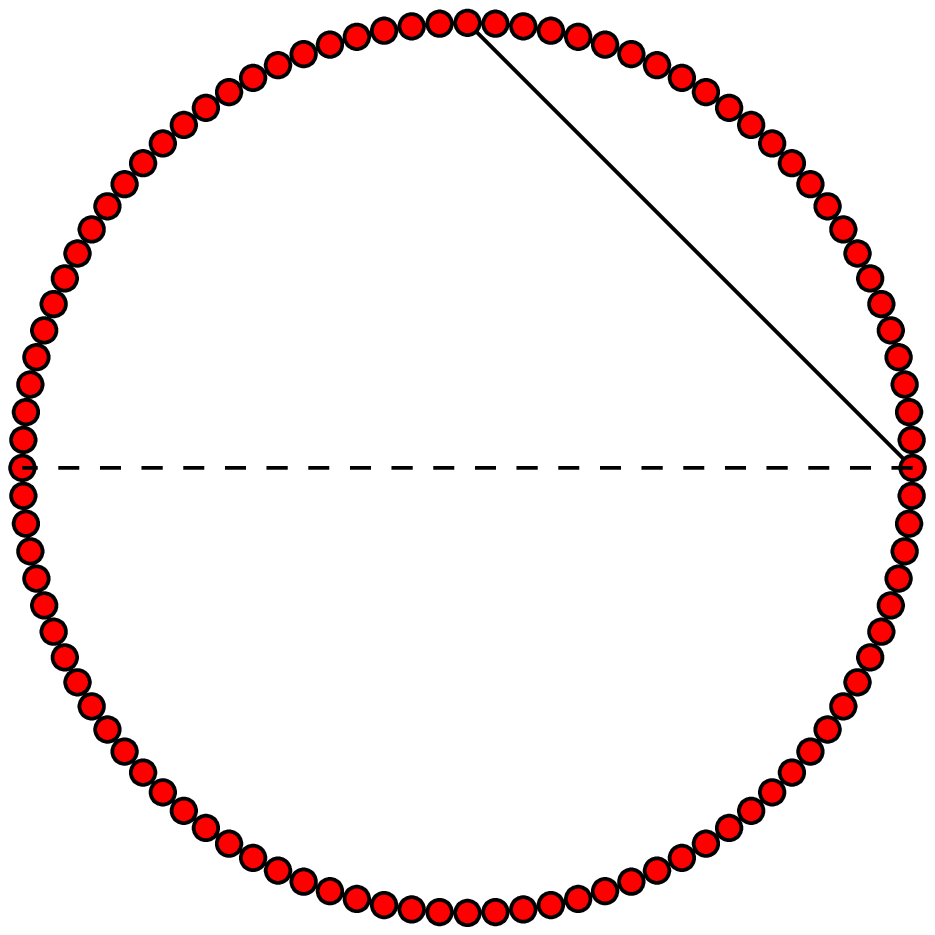}
\label{offset25}
}
\caption{Changes on the position of the extra link for $C_{N,k}^+$ with $N=100$ nodes and $k=1$ (i.e., $p=0.01$), when (a) $\Delta=1.0$ and (b) $\Delta=0.49$. The black dashed line represents a guide for the eye to locate the extra-link of the case $\Delta=1$.}
\label{redes_offset}
\end{figure}

Given a $(N,k)-$cycle graph, let us define a $(N,k)-$\textit{modified cycle graph}, $C_{N,k}^+$, as the network that is obtained by adding one extra-link to $C_{N,k}$. To characterize the $C_{N,k}^+$ topology, we use a \textit{shortcut position parameter}, denoted by $\Delta$. It is defined as follows:

\begin{equation}
\Delta =\frac{\theta_{ij}}{\pi}\textnormal{, for $\frac{2\pi}{N}k<\theta_{ij}$}
\label{def_Delta}
\end{equation}

\noindent where nodes $i$ and $j$ are the shortcut ends (i.e., $d_{ij}>1$), $\theta_{ij}$ is central angle (of the minor circle sector) defined by $i$ and $j$ when the $N$ nodes are uniformly distributed in an unit circle. That is, $\Delta$ is a non-negative number that we use to describe how far the ends of the extra-link are. Thus, the larger the value of $\Delta$, the further away the shortcut ends. Note that according to Eq.~\ref{def_Delta}, when $\Delta\leq 2k/N$, we obtain a $C_{N,k}$, since $C_{N,k}$'s and $C_{N,k}^+$'s are simple graphs. On the other hand, we match the case of $\Delta=1$ (i.e., $d_{i,j}=d_{\mathrm{max}}$) to a $C_{N,k}^+$ graph that is equidivided by a shortcut (see Fig.~\subref*{offset0} for an example). For this reason, hereafter, we only consider $C_{N,k}^+$ graphs with an even number of nodes.

The graphs described previously (i.e., $(N,k)-$cycles and modified cycles) are particular cases of the NW-model \cite{Newman99, Newman_S_W_01}, which is obtained by starting from a $C_{N,k}$ arrangement and following just one simple rule: for each existing edge $i-j$ a new link $i-k$ (called shortcut), with a randomly-chosen node $k$ is added with probability $p$. Consequently, the probability that a given site has a shortcut is given by $kp$, and, on average, there will be $x=kpN$ shortcuts in the NW-arrangements. Its easy to see that $C_{N,k}$ and $C_{N,k}^+$ graphs relate to NW-networks with $x=0$ and $x=1$, respectively.

\begin{figure}[h!]
\centering
\includegraphics[width=.5\textwidth]{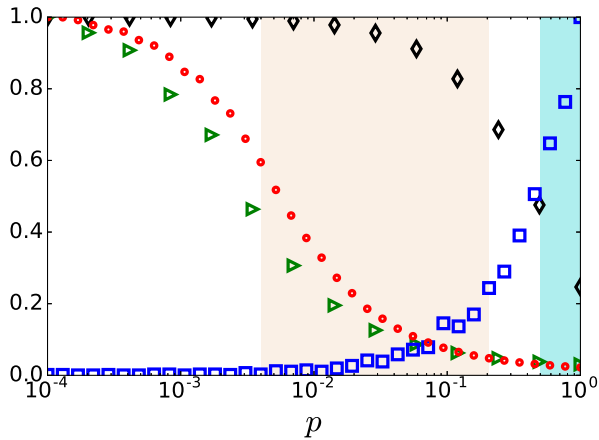}
\caption{Numerical results (averaged over 500 independent realizations) for the dependence of characteristic path length $L(p)$ and of clustering coefficient $C(p)$ on $p$ for a  NW-network with $N=1000$ nodes, when $k=1$ and 2, respectively. Results for $k=1$: $C(p)/C(1)$ (blue squares) and $L(p)/L(0)$ (red dots). Results for $k=2$: $C(p)/C(0)$ (black diamonds) and $L(p)/L(0)$ (green triangles).}
\label{SW_modelos}
\end{figure}

The NW-model is a slight variation on the well-known WS-model \cite{watts98} and, when $k>1$, the dependence of the small-world behavior of both models on $p$ is similar in the limit of large system sizes. That is, for intermediate values of $p$, the graphs are a small-world networks: with some small patches with reminiscent regular ordering linked by increasing number random links responsible for reducing the average shortest path lengths between nodes (see Fig.~\ref{SW_modelos} for an example, and Refs.~\cite{watts98,porter12} for further details). However, when $k=1$, the previous behavior varies. In an initial N-ring topology, the larger the value of $p$ (i.e., the amount of extra-links), the larger the average clustering coefficient (see Fig.~\ref{SW_modelos}). In order to compare our findings with previous analytical results for NW-arrangements with $k=1$ (see Refs.~\cite{kulkarni00,almaas02,almaas03}), in this work, we also consider the sparse regime $p\ll1$.

\section{Markov chain results}
\label{results}

In this section we present MC results for the time evolution of $\left \langle r^2(t) \right \rangle$ on NW-networks with $x=0$, 1 and $x>1$, with the condition that $p\ll1$. As discussed in Sec.~\ref{MSD_markov}, our results are limited to $N\leq1500$.

\begin{figure}[h!]
\centering
\subfloat[]{
   \begin{tikzpicture}
        \node[anchor=south west,inner sep=0] (image) at (0,0) {\includegraphics[width=0.5\textwidth]{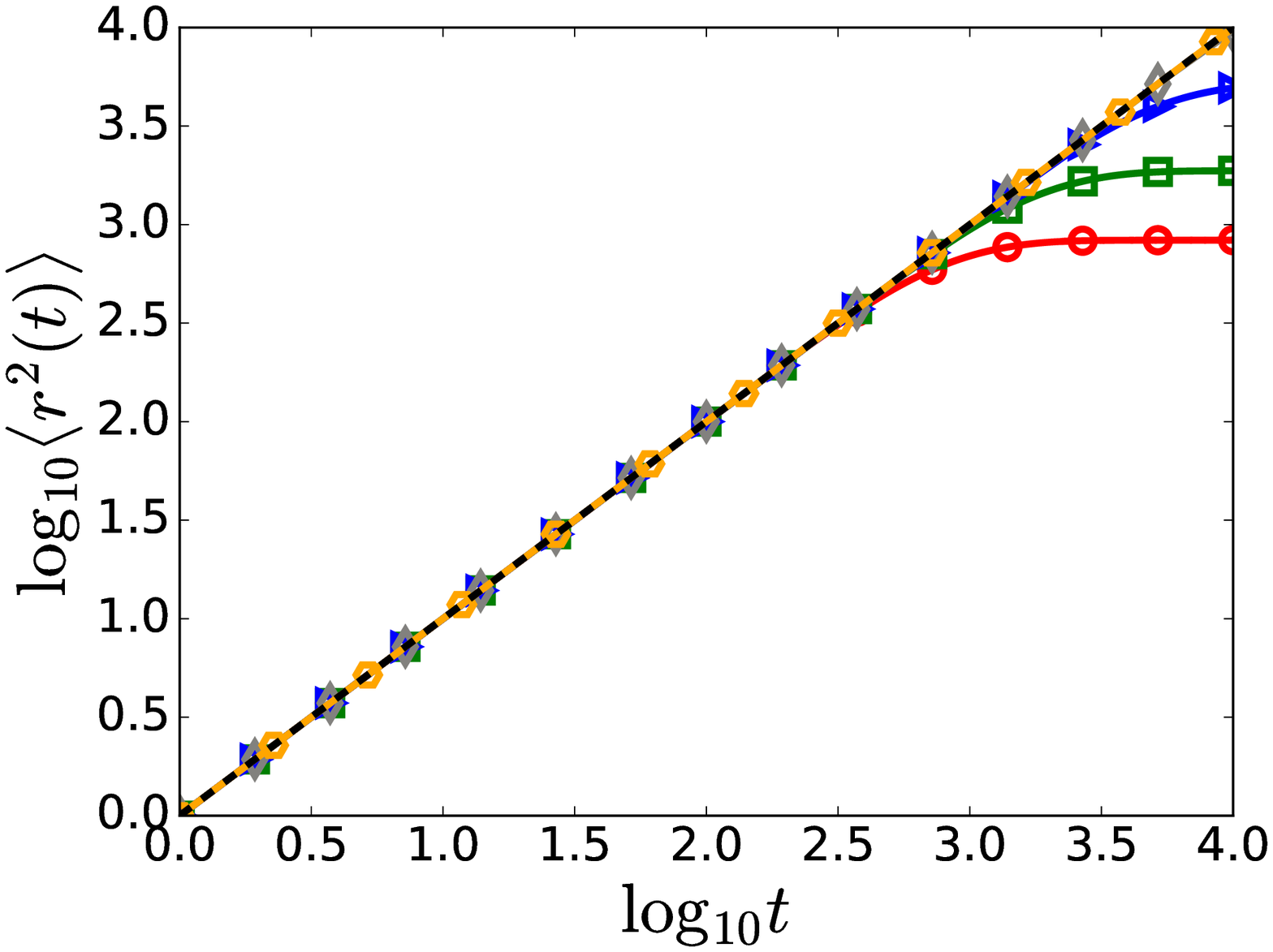}};
        \begin{scope}[x={(image.south east)},y={(image.north west)}]
            \node[anchor=south west,inner sep=0] (image) at (0.19,0.56) {\includegraphics[width=0.175\textwidth]{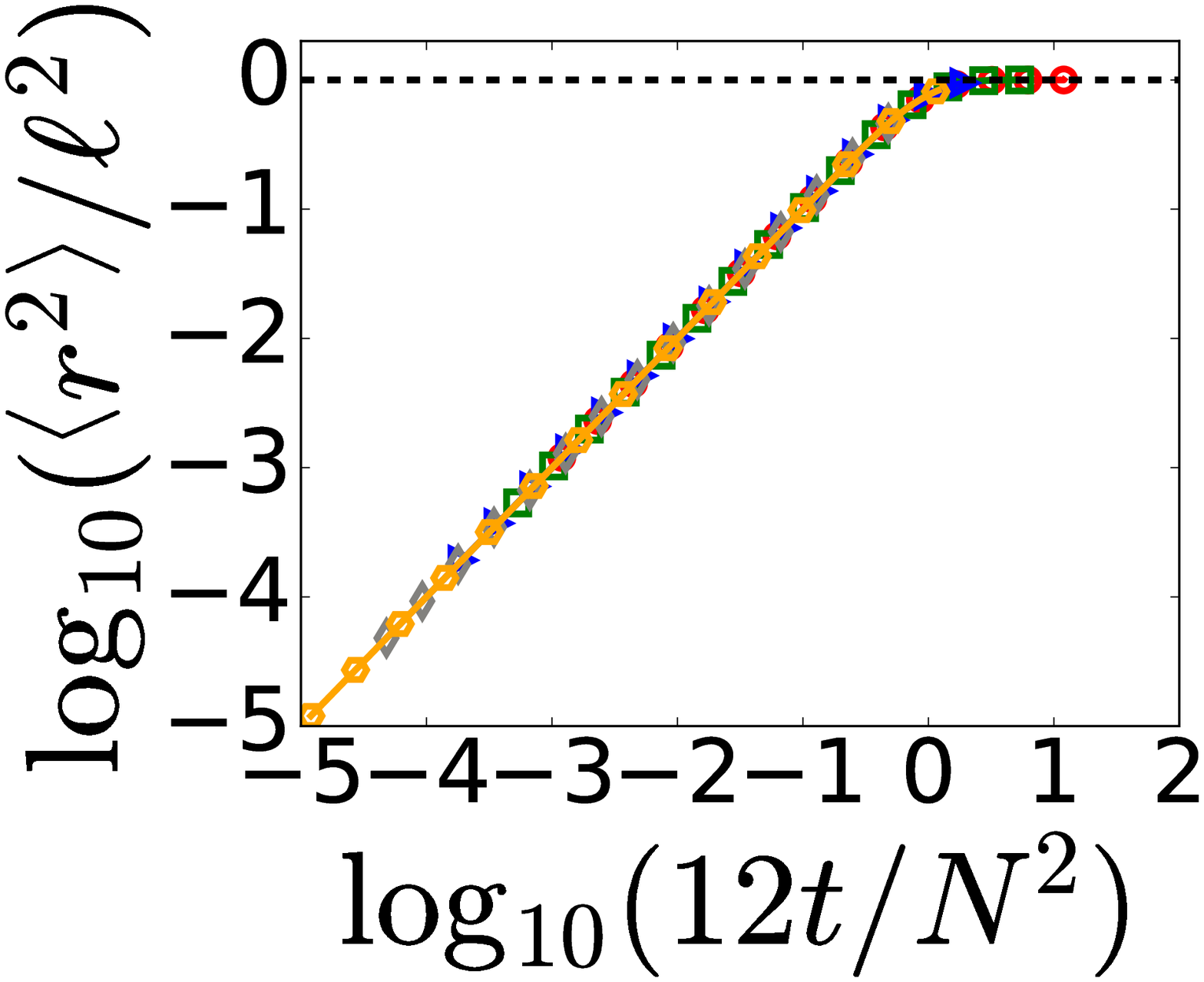}};
        \end{scope}
    \end{tikzpicture}
\label{MSD_cyclos_k1}
}
\subfloat[]{
   \begin{tikzpicture}
        \node[anchor=south west,inner sep=0] (image) at (0,0) {\includegraphics[width=0.5\textwidth]{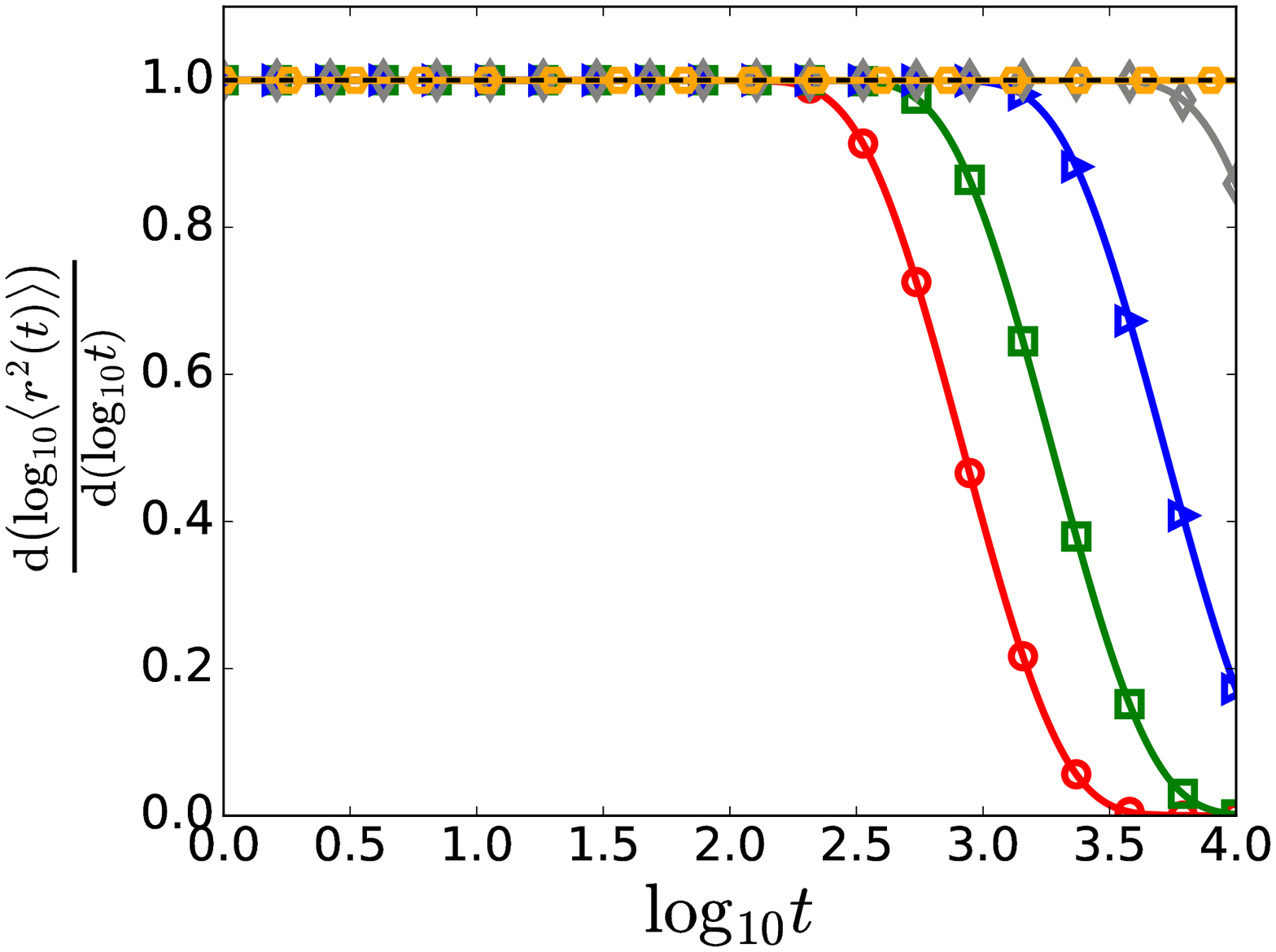}};
        \begin{scope}[x={(image.south east)},y={(image.north west)}]
            \node[anchor=south west,inner sep=0] (image) at (0.22,0.18) {\includegraphics[width=0.20\textwidth]{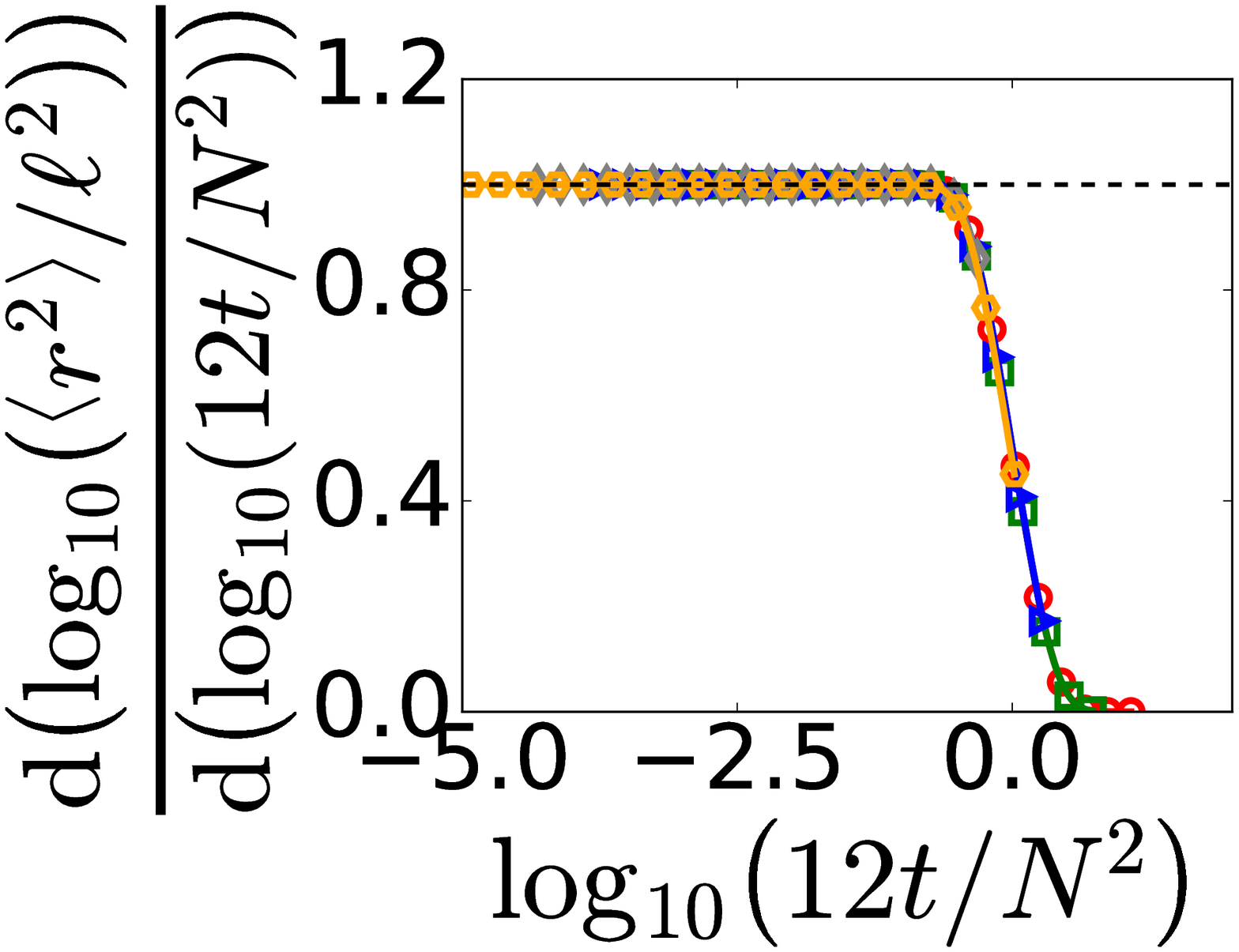}};
        \end{scope}
    \end{tikzpicture}
\label{Der_MSD_cyclos_k1}
}
\caption{(a) Time evolution of $\left \langle r^2(t) \right \rangle$ on cycle graphs with $k=1$: $N=100$ (red circles), $N=150$ (green squares), $N=250$ (blue triangles), and $N=500$ (grey diamonds) and $N=1000$ (orange hexagons). (b) Numerical derivative of $\log_{10} \left \langle r^2(t) \right \rangle$ with respect to $\log_{10}t$ obtained for the series shown in Fig.~\ref{cycles_k1}(a). The insets show the data collapses of the curves obtained from Eq.~(\ref{approx_cycle}). The black dashed lines in both panels are a guide for the eye to locate a normal diffusion.}
\label{cycles_k1}
\end{figure}

\subsection{NW-networks with $x=0$}
\label{sub_sec_x0}

For the sake of comparison, we first consider the simple case of finite $(N,k)-$cycle graphs (i.e., $p=0$). In Fig.~\ref{cycles_k1}, we show results for $k=1$. As expected, these systems exhibit a normal diffusion (i.e., $\left \langle r^2(t) \right \rangle \sim t^\gamma$ with $\gamma=1$), before finite size effects (or saturation) take place [see Fig.~\subref*{MSD_cyclos_k1}]. According to those features, the time evolution can be summarized as:

\begin{eqnarray}
\left \langle r^2(t) \right \rangle = \left\{ \begin{array}{r}
t,\\
\ell^2,
\end{array}\right.\begin{array}{l}
\textnormal{if \ensuremath{t\ll N^2/12}}\\
\textnormal{if \ensuremath{t\gg  N^2/12}}
\end{array},
\label{approx_cycle}
\end{eqnarray}

\noindent where $\ell^2$ is the squared minimum distance between a pair of nodes, which for simple undirected networks is given by:

\begin{equation}
\ell^2=\frac{1}{N^2} \sum_{i=1}^N \sum_{j=1}^N  \left(  \hat{\mathbf{M}}\circ \hat{\mathbf{M}}\right )_{ij},
\label{squared_dist}
\end{equation}

\noindent where $\hat{\mathbf{M}}$ denotes the neighborhood matrix, as defined in Ref.~\cite{andrade06}, $\circ$ represents the Hadamard product, and $(\mathbf{X})_{ij}$ is the element $ij$ of matrix $\mathbf{X}$. In the case of $(N,k)-$cycle graphs, it is possible to express $\ell^2$ as:

\begin{equation}
\ell^2=\frac{1}{N}\left ( -\frac{4}{3}k d_{\mathrm{max}}^3+\left ( N-1+k \right )d_{\mathrm{max}}^2+\frac{k}{3}d_{\mathrm{max}} \right ),
\label{squared_dist_cycle}
\end{equation}

\noindent where

\begin{equation}
d_{\mathrm{max}}=\left \lceil \frac{N}{2k} \right \rceil,
\label{diameter_cycle}
\end{equation}

\noindent and $\left \lceil . \right \rceil$ represents the ceiling function. In the insets of Fig.~\ref{cycles_k1}, we show the excellent scaling collapses obtained for $k=1$ by using Eq.~(\ref{approx_cycle}), confirming that the expected saturation value $\ell^2 = (N^2+2)/12 \approx N^2/12$  in Eq.~(\ref{squared_dist_cycle}) is obtained at large values of $t$. We also confirm the results for $\gamma$ by estimating the numerical Log-derivatives of the prior series [see Fig.~\subref*{Der_MSD_cyclos_k1}]. Once the MC formalism only provides $\left \langle r^2(t) \right \rangle$ values for discrete time-steps, the derivatives were obtained by considering values at neighboring integer values of $t$.

\begin{figure}[h!]
\centering
\subfloat[]{
   \begin{tikzpicture}
        \node[anchor=south west,inner sep=0] (image) at (0,0) {\includegraphics[width=0.5\textwidth]{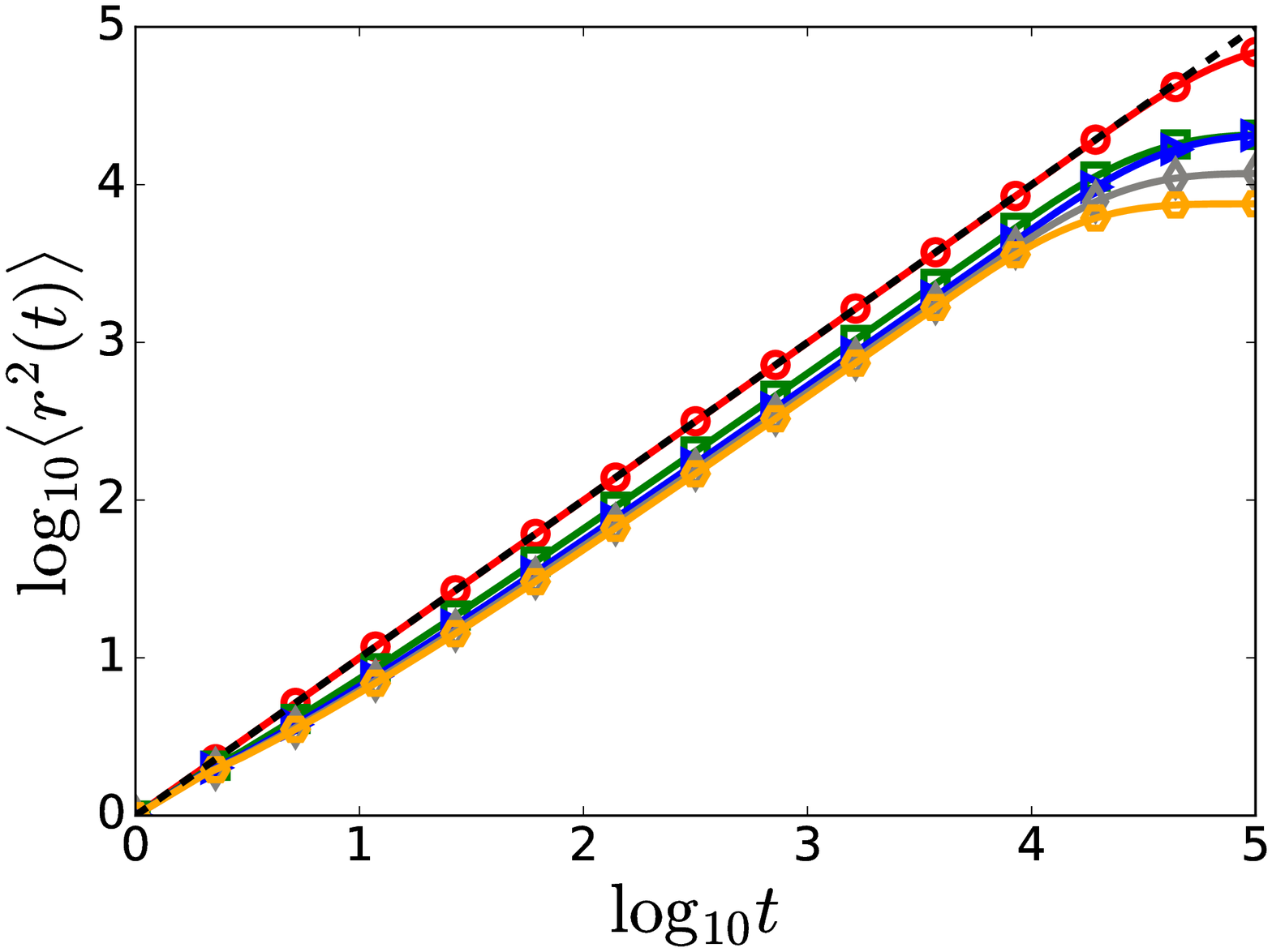}};
        \begin{scope}[x={(image.south east)},y={(image.north west)}]
            \node[anchor=south west,inner sep=0] (image) at (0.155,0.54) {\includegraphics[width=0.185\textwidth]{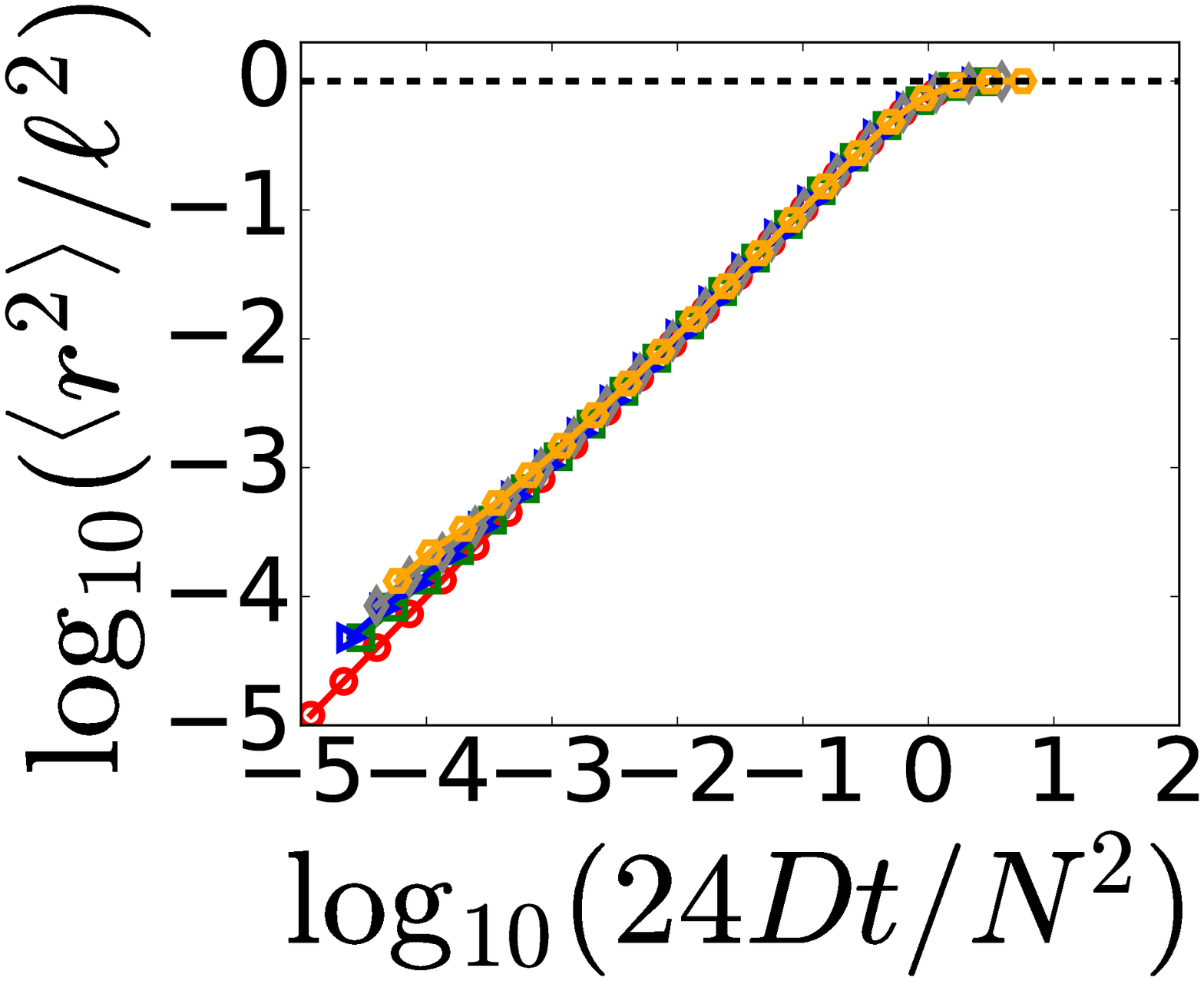}};
        \end{scope}
    \end{tikzpicture}
\label{MSD_cyclo_k2}
}
\subfloat[]{
   \begin{tikzpicture}
        \node[anchor=south west,inner sep=0] (image) at (0,0) {\includegraphics[width=0.5\textwidth]{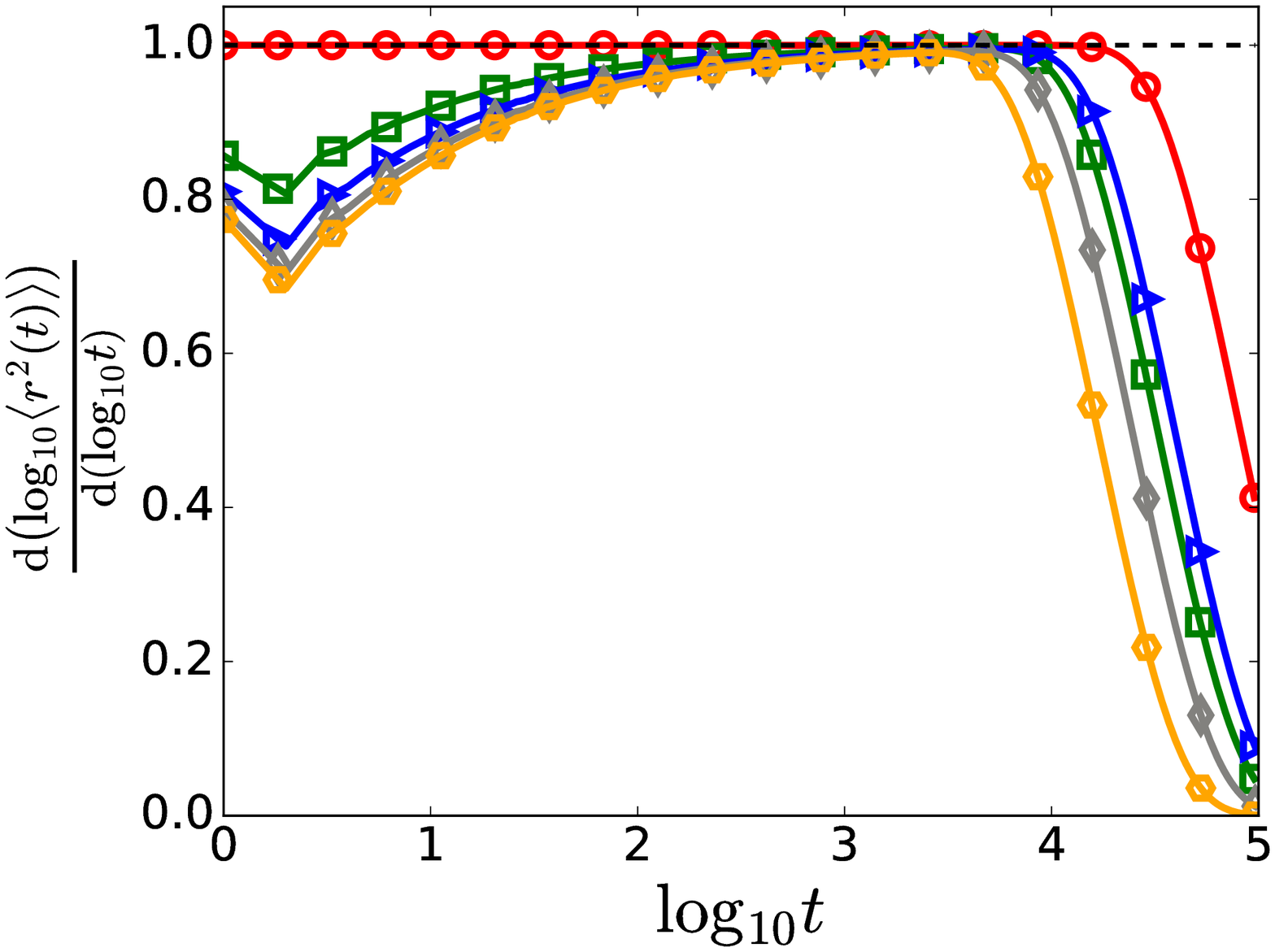}};
        \begin{scope}[x={(image.south east)},y={(image.north west)}]
            \node[anchor=south west,inner sep=0] (image) at (0.20,0.165) {\includegraphics[width=0.258\textwidth]{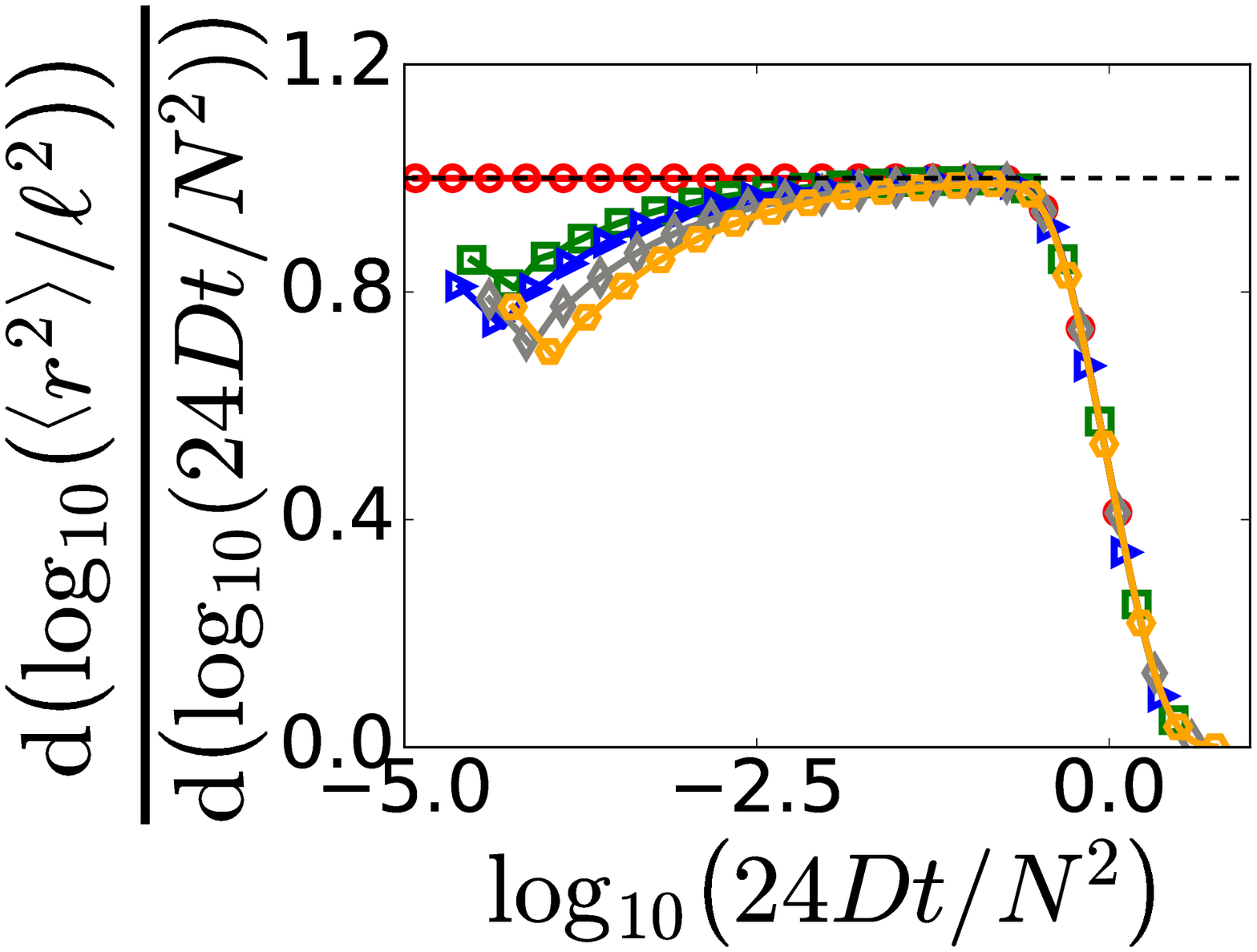}};
        \end{scope}
    \end{tikzpicture}
\label{der_MSD_cyclo_k2}
}
\caption{(a) Time evolution of $\left \langle r^2(t) \right \rangle$ on $(N,k)-$cycle graphs with distinct $(N,k)$ combinations: $(1000,1)$ (red circles), $(1000,2)$ (green squares), $(1500,3)$ (blue triangles), $(1500,4)$ (grey diamonds), and $(1500,5)$ (orange hexagons). (b) Numerical derivative of $\log_{10} \left \langle r^2(t) \right \rangle$ with respect to $\log_{10}t$ obtained for the series in Fig.~\ref{cycles_k2}(a). The insets show the data collapse obtained from Eq.~(\ref{approx_cycle_k2}) for the Gaussian and saturation regimes of $\left \langle r^2(t) \right \rangle$ curves and their Log-derivatives. The black dashed lines in both panels are a guide for the eye to locate a normal diffusion.}
\label{cycles_k2}
\end{figure}

Figure~\ref{cycles_k2} illustrates the behavior of $\left \langle r^2(t) \right \rangle$ when $k>1$ and $N$ is relatively large, with the presence of a transient sub-diffusive regime during the first time-steps that does not appear when $k=1$. For large $N$, the emergence of a normal diffusive behavior for large $t$ indicates that the central limit theorem (CLT) holds for random walks on $(N,k)-$cycle graphs, before finite size effects manifest \cite{Masuda17}. The Log-derivatives show that the larger the number of direct neighbors per node (i.e., $2k$), the slower the convergence to the normal behavior [see Fig.~\subref*{der_MSD_cyclo_k2}]. On the other hand, we can see that the larger the value of $k$, the smaller the saturation value. The addition of direct-neighbors reduces the shortest path distances between the nodes of the network (i.e., the elements of $(\hat{\mathbf{M}})_{ij}$), and, therefore, the corresponding value of $\ell^2$ [see  Eqs.~(\ref{squared_dist}) and (\ref{squared_dist_cycle})]. As a consequence, in the case of small $(N,k)-$cycle graphs with $k>1$, the transitory subdiffusive regime may overlap with their saturation process. It is worth mentioning that this induced subdiffusion is caused by finite size effects, not by a violation of the CLT. For large enough $(N,k)-$cycle graphs with $k\geq1$, the Gaussian and saturation regimes of $\left \langle r^2(t) \right \rangle$ can be described as:

\begin{eqnarray}
\left \langle r^2(t) \right \rangle = \left\{ \begin{array}{r}
\sum _{j=1}^{kt}2\left ( \left \lceil \frac{j}{k} \right \rceil \right )^2 P(j,t),\\
\ell^2,
\end{array}\right.\begin{array}{l}
\textnormal{for \ensuremath{10\lesssim t\ll N^2/(4k^2+6k+2)}}\\
\textnormal{for \ensuremath{t\gg  N^2/(4k^2+6k+2)}}
\end{array},
\label{approx_cycle_k2}
\end{eqnarray}

\noindent where

\begin{equation}
P(j,t)=\left ( \pi 4 Dt \right )^{-1/2}\exp\left ( \frac{-j^2}{4Dt} \right )=\left ( \pi \frac{2k^2+3k+1}{3}t \right )^{-1/2}\exp\left ( \frac{-3j^2}{(2k^2+3k+1)t} \right )
\label{Prob_inf_cycle}
\end{equation}

\noindent represents the probability of finding the random walker at node $j\in \mathbb{Z}$ (\enquote{diffusion front}), after performing $t$ steps, in the thermodynamic limit ($N\rightarrow \infty$), when the walker is initially located at node $j=0$ and $t$ is large (see Appendix). By setting $k=1$ in Eq.~(\ref{approx_cycle_k2}), we recover the behavior described in Eq.~(\ref{approx_cycle}) for $t\gtrsim10$. In the insets of Fig.~\ref{cycles_k2}, we show the data collapse obtained for different $(N,k)$ combinations, by Eq.~(\ref{approx_cycle_k2}).


\begin{figure}[h!]
\centering
\subfloat[]{
\includegraphics[width=0.5\linewidth]{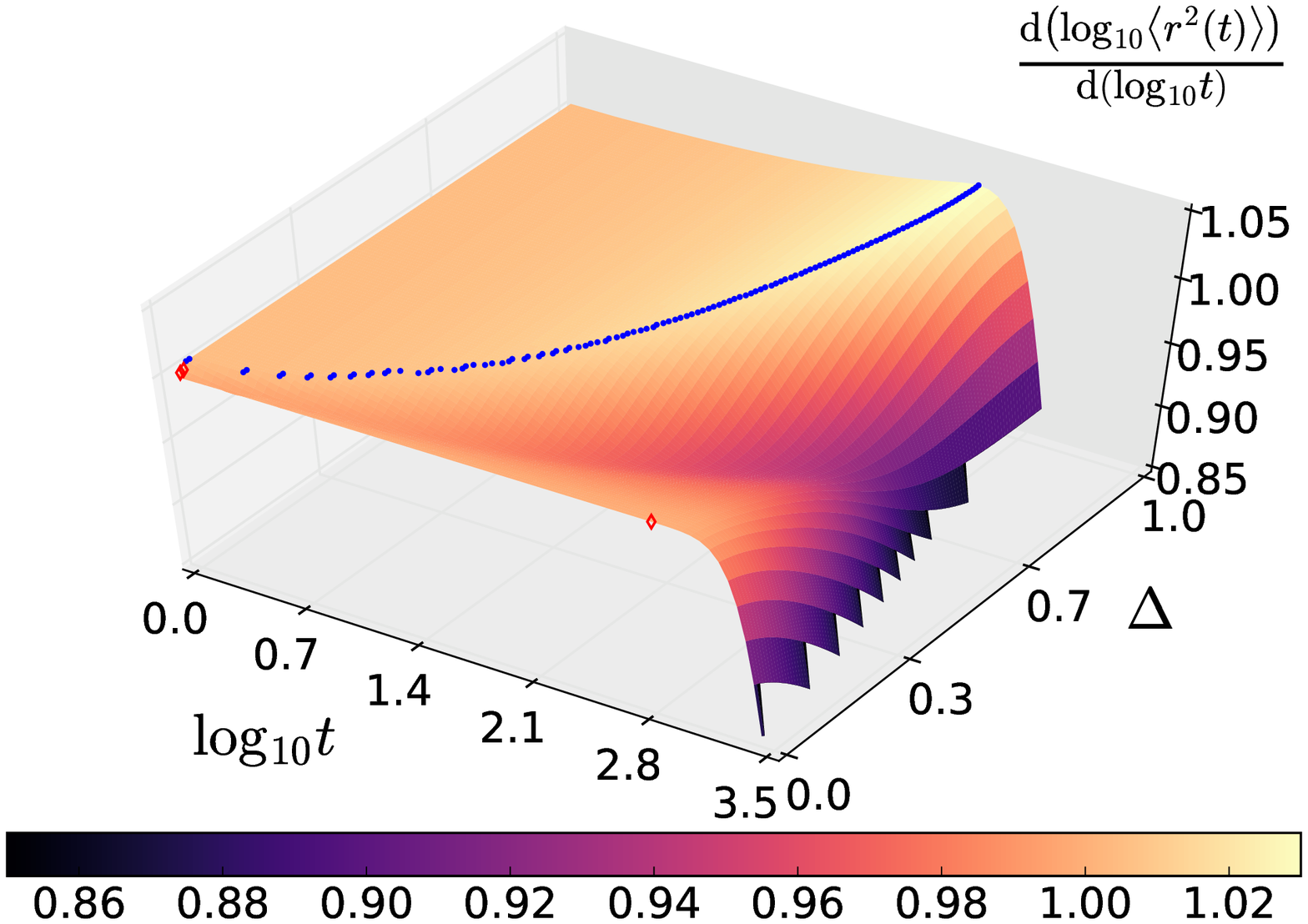}
\label{3D_250_k1}
}
\subfloat[]{
\includegraphics[width=0.5\linewidth]{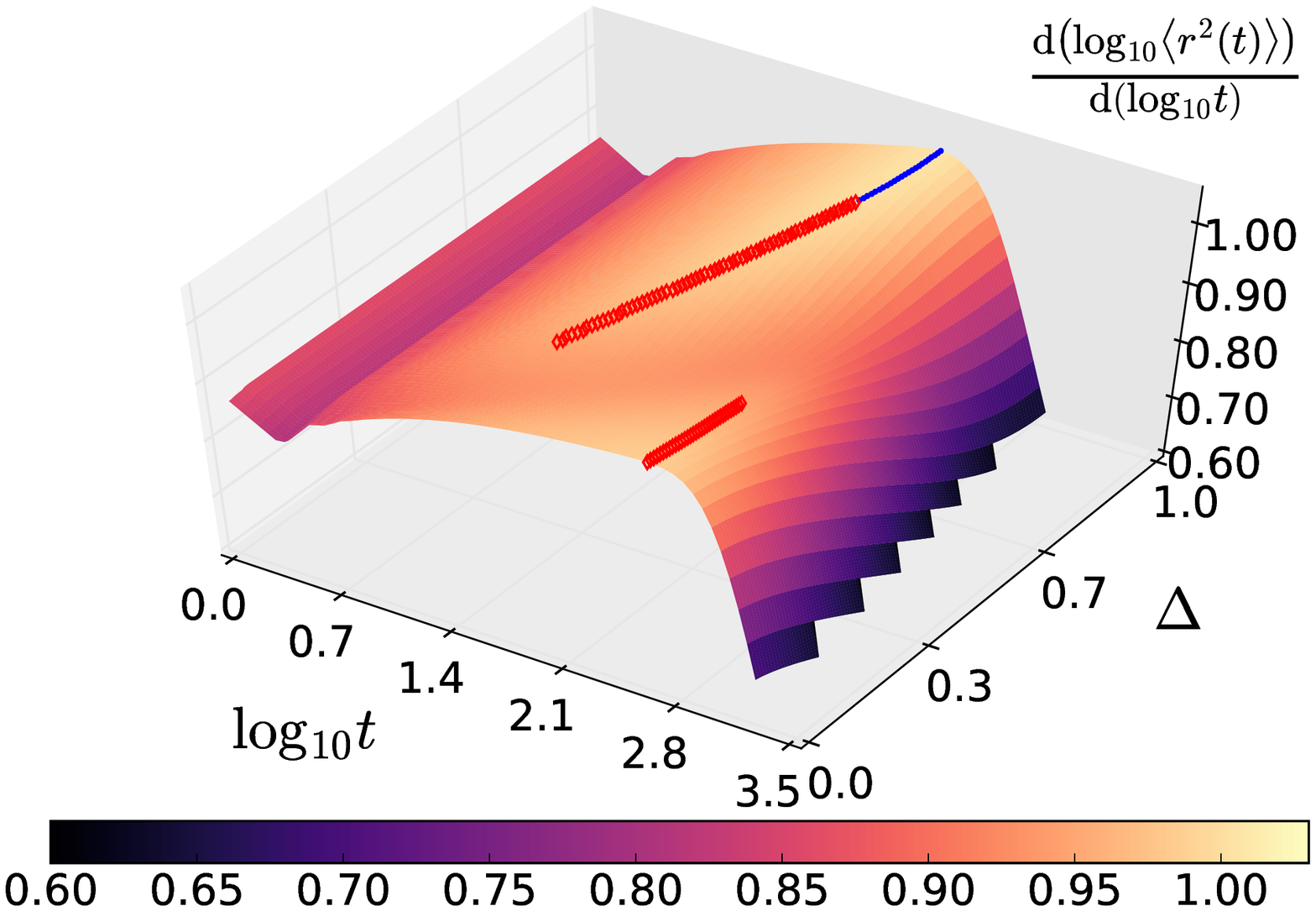}
\label{3D_250_k2}
}
\caption{Numerical derivative of $\log_{10} \left \langle r^2(t) \right \rangle$ with respect to $\log_{10}t$ for $C_{N,k}^+$ graphs with $N=250$ nodes, when $0<\Delta\leq1$: (a) $k=1$ and (b) $k=2$. Superdiffusive results are indicated with  blue dots (at the maximum of the respective Log-derivative), while subdiffusive ones are ploted with red diamonds.}
\label{3D_250}
\end{figure}

\subsection{NW-networks with $x=1$}
\label{subsec_x1}

In this Subsection we study the average evolution of $\left \langle r^2(t) \right \rangle$ on $C_{N,k}^+$ graphs, i.e., NW-networks with $x = 1$. It is worth mentioning that, depending on the chosen values of $N$, $k$ and $\Delta$, different diffusive patterns can be observed. 
Our MC calculations show that the larger the value of $\Delta$, the faster the diffusion. In Fig.~\ref{3D_250}(a) and (b), we present the dependence of the Log-derivatives of $\left \langle r^2(t) \right \rangle$ on $t$ and $\Delta$ for $N=250$, when $k=1$ and $k=2$. The intesection of the surfaces with a plane of $\Delta=$constant produces curves that are similar to those shown in Fig.~\subref*{der_MSD_cyclo_k2}. For a given value of $\Delta$, the position of the absolute maximum of the Log-derivative of $\left \langle r^2(t) \right \rangle$ with respect to time is indicated by a blue dot or a red diamond according to whether its value is greather than 1 or not, respectively. As can be observed, the larger the value of $\Delta$, the faster the diffusion independently of the value of $k$. On the other hand, for a given size $N$, the larger the value of $k$, the larger the number of values of $\Delta$ with sub-diffusive behavior. The findings for $k=1$ suggest that subdiffusion is caused by the creation of a small isolated cluster or \enquote{subcycle} by the extra-link and the $(N,k)-$cycle structure. Such \enquote{subcycle} may trap the random walker inside it and, thus, hinder its diffusion. On the other hand, when $k>1$, we observe subdiffusion for every $\Delta$ during the first time-steps. As in the case of $(N,k)$-cycles, $C_{N,k}^+$ graphs also exhibit a transitory subdiffusive regime for $k>1$.

\begin{figure}[h!]
\centering
\subfloat[]{
   \begin{tikzpicture}
        \node[anchor=south west,inner sep=0] (image) at (0,0) {\includegraphics[width=0.5\textwidth]{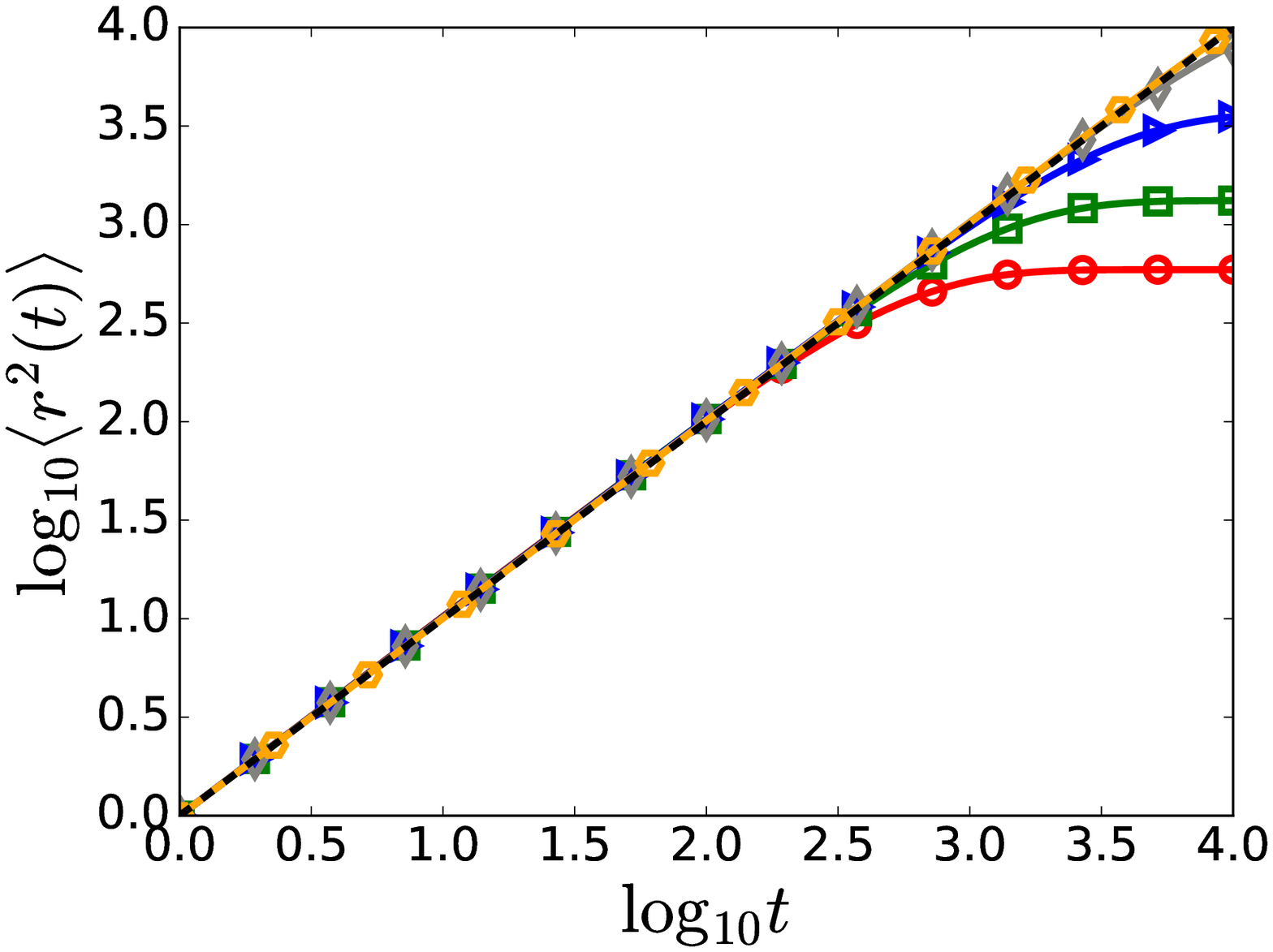}};
        \begin{scope}[x={(image.south east)},y={(image.north west)}]
            \node[anchor=south west,inner sep=0] (image) at (0.165,0.535) {\includegraphics[width=0.182\textwidth]{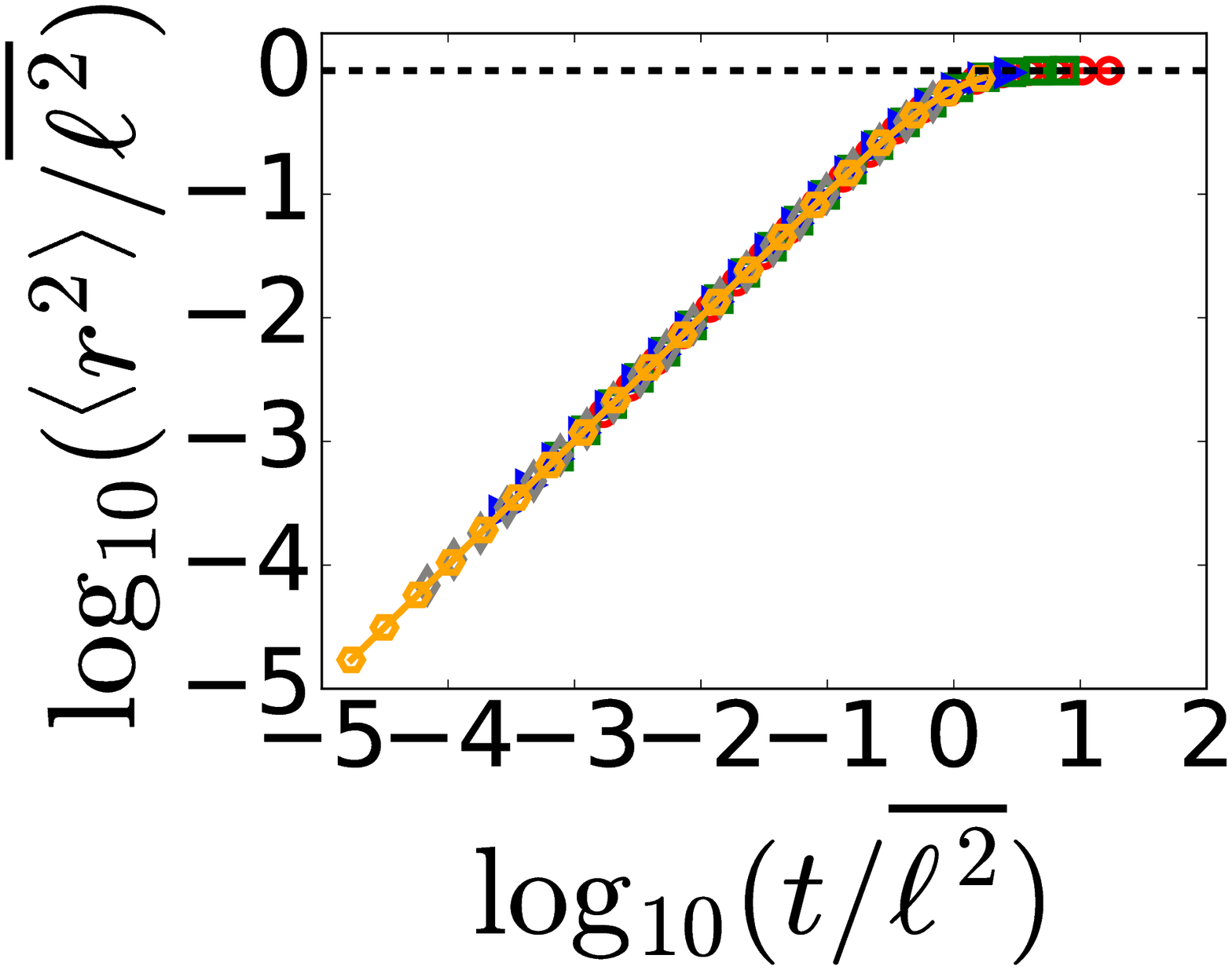}};
        \end{scope}
    \end{tikzpicture}
\label{MSD_SUMA}
}
\subfloat[]{
   \begin{tikzpicture}
        \node[anchor=south west,inner sep=0] (image) at (0,0) {\includegraphics[width=0.5\textwidth]{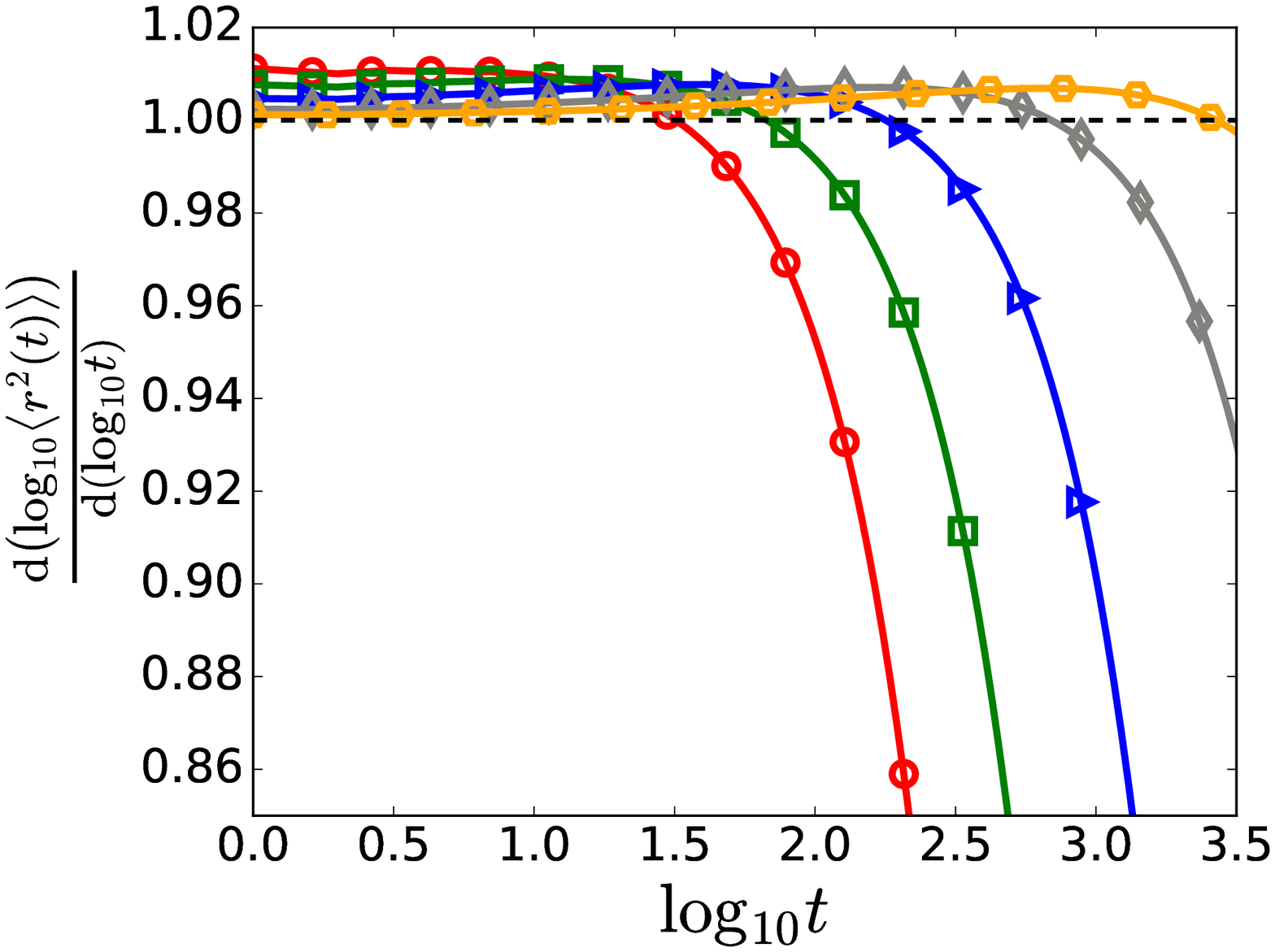}};
        \begin{scope}[x={(image.south east)},y={(image.north west)}]
            \node[anchor=south west,inner sep=0] (image) at (0.22,0.18) {\includegraphics[width=0.19\textwidth]{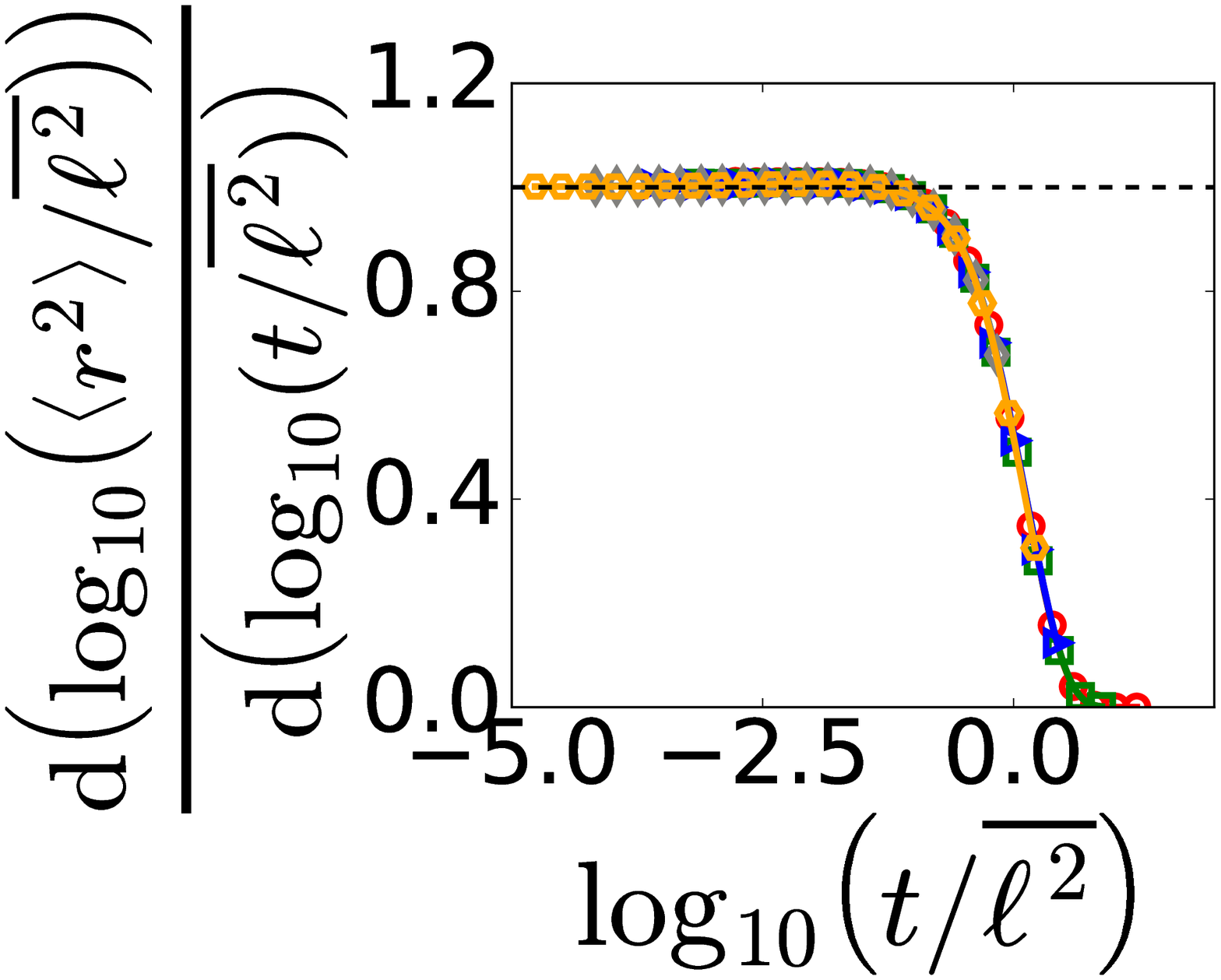}};
        \end{scope}
    \end{tikzpicture}
\label{der_MSD_SUMA}
}
\caption{(a) Dependence of the time evolution of $\left \langle r^2(t) \right \rangle$ for $C_{N,k}^+$ graphs (averaged over $\Delta$): $N=100$ (red circles), $N=150$ (green squares), $N=250$ (blue triangles), $N=500$ (grey diamonds), and $N=1000$ (orange hexagons). (b) Detail of the numerical derivative of $\log_{10} \left \langle r^2(t) \right \rangle$ with respect to $\log_{10}t$ for the series in Fig.~\ref{MSD_MCGs_SUMA}(a). The insets show the data collapses obtained using the scaling ansatz in Eq.~(\ref{scaling_k1_x1}). The black dashed line is a guide for the eye to locate a normal diffusion.}
\label{MSD_MCGs_SUMA}
\end{figure}

Regarding the behavior of $\Delta$ averaged $C_{N,k}^+$ graphs, in Fig.~\ref{MSD_MCGs_SUMA} we present the time evolution of $\left \langle r^2(t) \right \rangle$ and the Log-derivatives for $k=1$ and various $N$. According to the Log-derivatives, the diffusion exponent $\gamma$ is greater than one, before saturation appears [see Fig.~\subref*{der_MSD_SUMA}]. Therefore, in the case of $k=1$, it is possible to identify a superdiffusive behavior. However, given that $\gamma$ is very close to 1, it is possible to obtain excellent data collapse from $\left \langle r^2(t) \right \rangle$ curves (see the insets in Fig.~\ref{MSD_MCGs_SUMA}) by using the following scaling ansatz:

\begin{equation}
 \left \langle r^2(t) \right \rangle \approx \left\{ \begin{array}{r}
t,\\
\overline{\ell^2},
\end{array}\right.\begin{array}{l}
\textnormal{for \ensuremath{t\ll\overline{\ell^2}}}\\
\textnormal{for \ensuremath{t\gtrsim \overline{\ell^2} }}
\end{array},
\label{scaling_k1_x1}
\end{equation}

\noindent where $\overline{\ell^2}$ is the average squared minimum distance between a pair of nodes. Note that we do not use the scaling ansatz proposed in Refs.~\cite{almaas02,almaas03} because, when $x=1$ and $k=1$, the condition $\xi<\sqrt{\overline{\ell^2}}$ is not met (see Subsec.~\ref{Compa_Sec}), where $\xi=1/p$ represents the average distance the walker travels to reach a shortcut in Refs.~\cite{Newman99,almaas03}.

\begin{figure}[h!]
\centering
\subfloat[]{
\includegraphics[width=0.5\linewidth]{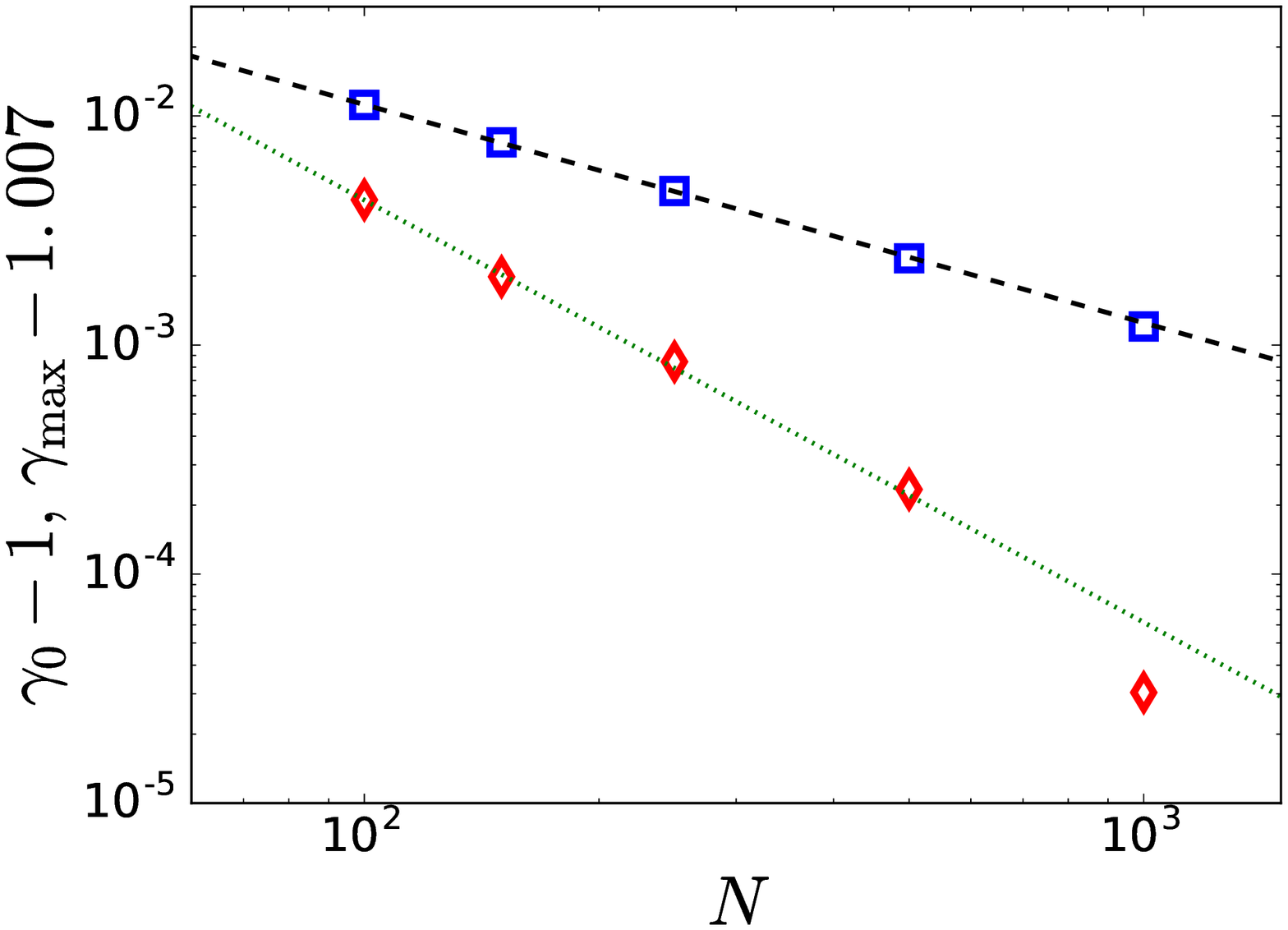}
\label{gamma_SUMA_k1}
}
\subfloat[]{
\includegraphics[width=0.5\linewidth]{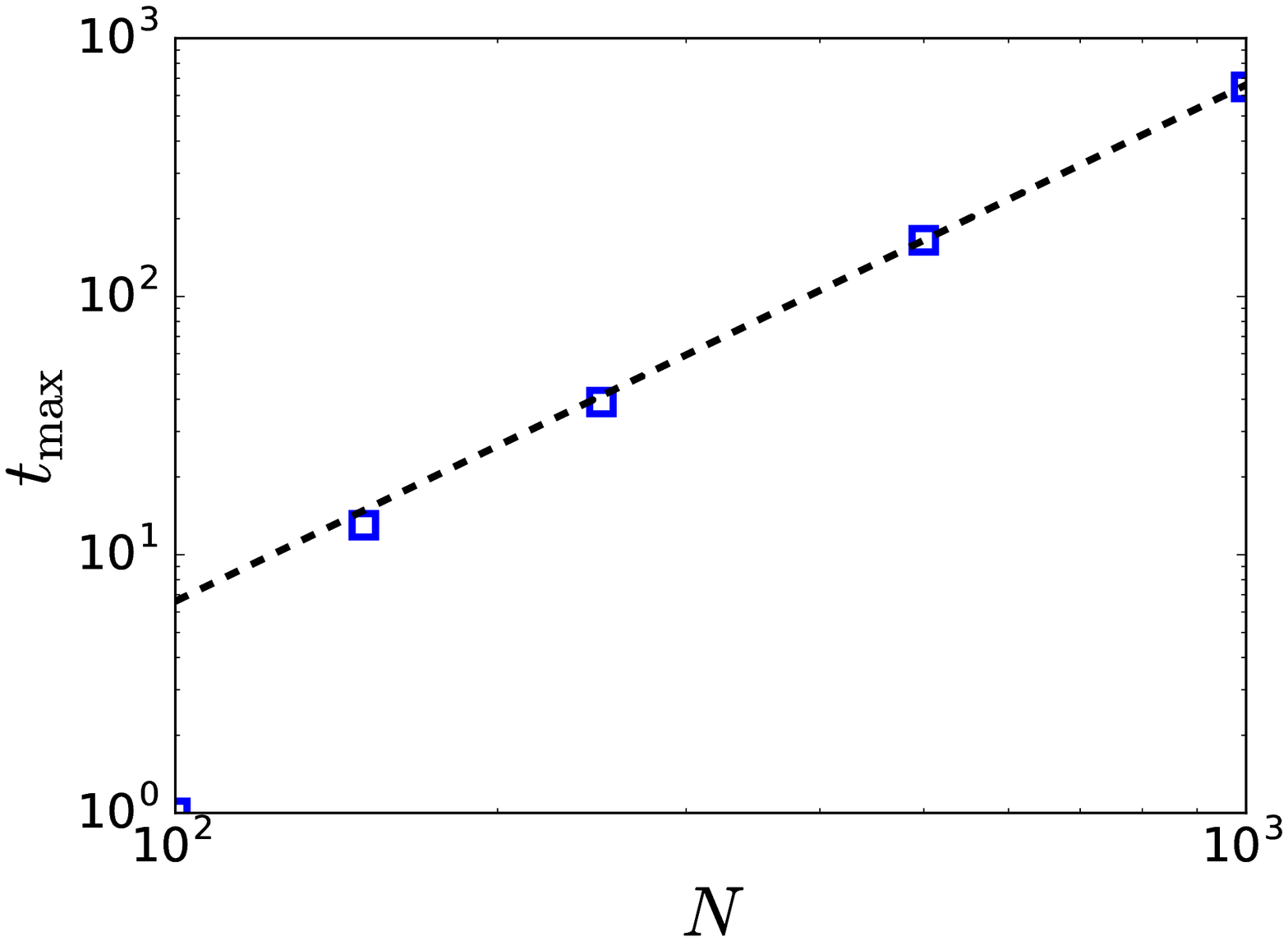}
\label{duration_SUMA_k1}
}
\caption{(a) Dependence of $\gamma_0-1$ (blue squares) and  $\gamma_{\mathrm{max}}-1.007$ (red diamonds) on $N$ for $\Delta$ averaged $C_{N,k}^+$ graphs. The black dashed line represents adjusted dependence of $\gamma_{0}$ on $N$, given by $\gamma_{0}-1\approx 0.903N^{-0.953}$. The dotted line shows the adjusted dependence of $\gamma_{\mathrm{max}}$ on $N$, given by $\gamma_{\mathrm{max}}-1.007\approx 20.854N^{-1.843}$ (b): Dependence of $t_{\mathrm{max}}$ on $N$ for $\Delta$ averaged $MC_{N,k}$ graphs (blue dots). The black dashed line is a guide proportional to $N^2$.}
\label{duration_gamma_SUMA}
\end{figure}

On the other hand, we can see in Fig.~\subref*{der_MSD_SUMA} that, before saturation, $\gamma$ grows slightly from an initial value, $\gamma_0$, to a maximum one, $\gamma_{\mathrm{max}}$, that is reached at $t=t_{\mathrm{max}}$. As can be seen in  Fig.~\subref*{gamma_SUMA_k1}, the larger the system size $N$, the smaller the value of  $\gamma_{0}$ and $\gamma_{\mathrm{max}}$. Indeed, small  or medium (average) $C_{N,k}^+$ graphs exhibit initially a superdiffusive behavior, whereas large ones show normal diffusion ($\gamma_0\approx 1$ for $N\gtrsim10^3$) during the first time-steps. However, in the case of large systems, $\gamma_{\mathrm{max}}$ remains almost constant and slightly greater than one ($\gamma_{\mathrm{max}}\approx 1.007$) so that superdiffusion emerges only at $t\sim t_{\mathrm{max}}\sim N^{2}$ [see Fig.~\subref*{duration_SUMA_k1}], just before finite-size effects become important and saturation takes place. That is, superdiffusion appears when the walker begins to reach the shortcut at average distance $\sqrt{\overline{\ell^2}}$ for $x=1$.

\begin{figure}[h!]
\centering
\subfloat[]{
\includegraphics[width=0.5\linewidth]{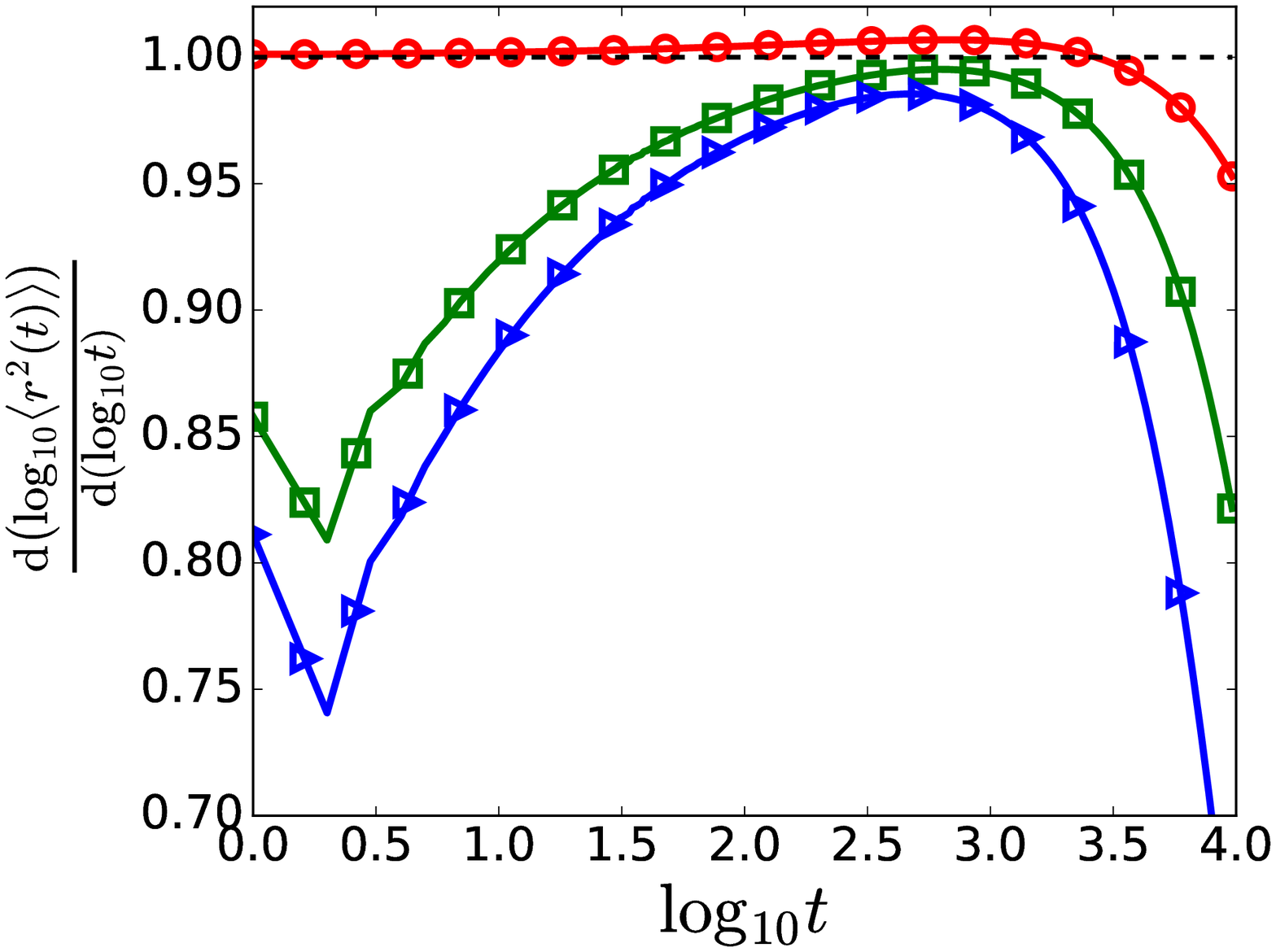}
\label{comp_x1_k1k2k3}
}
\subfloat[]{
\includegraphics[width=0.5\linewidth]{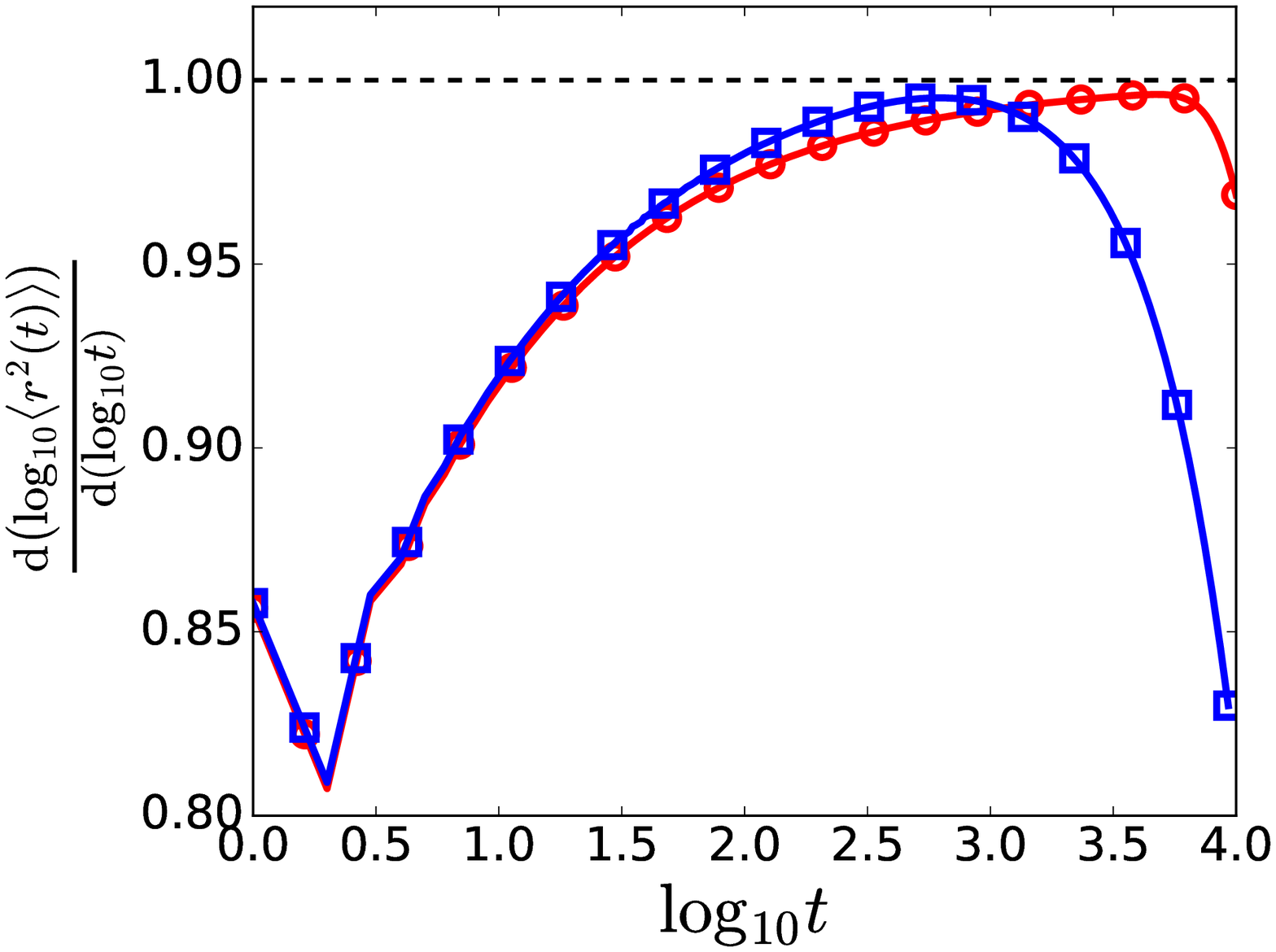}
\label{comp_x1_k1k2}
}
\caption{Detail of the numerical derivative of $\log_{10} \left \langle r^2(t) \right \rangle$ with respect to $\log_{10}t$ for an average NW-network with $N=1000$. (a) Results for $x=1$, when $k=1$ (red circles), $k=2$ (green squares) and $k=3$ (blue triangles). The black dashed line is a guide for the eye to locate a normal diffusion. (b) Results for $x=0$ (red circles) and $x=1$ (blue squares), when $k=2$. }
\label{comp_NW_x1_k1k2}
\end{figure}

Finally, in the case of average $C_{N,k}^+$ graphs with $k>1$, our calculations show that superdiffusion is replaced by subdiffusion, during the first-time steps. As with $(N,k)-$cycle graphs, for a given $N$, the larger the value of $k>1$, the slower the diffusion is (i.e., the smaller $\left \langle r^2(t) \right \rangle$ is) and the faster the saturation appears [see Fig.~\subref*{comp_x1_k1k2k3}]. Indeed, when $N\lesssim1000$, subdiffusion overlaps with the saturation process. On the other hand, in Fig.~\subref*{comp_x1_k1k2}, we can see that the addition of the extra-link makes the diffusion faster in $C_{N,k}^+$'s than in $C_{N,k}$'s, before finite size effects manifest. However, as expected, saturation also appear sooner (i.e., $\overline{\ell^2}$ is smaller), due to the shortcut. 

\subsection{NW-networks with $x>1$}
\label{sub_sec_x_mayor_1}

In this Subsection, we discuss the results for the sparse NW-networks with $x>1$ and $p=x/(kN)\ll1$. For each $(p,N,k)$ combination, we considered 100 independent network samples, and obtained average values of $\left \langle r^2(t) \right \rangle$ within the MC approach.

\begin{figure}[h!]
\centering
\subfloat[]{
   \begin{tikzpicture}
        \node[anchor=south west,inner sep=0] (image) at (0,0) {\includegraphics[width=0.5\textwidth]{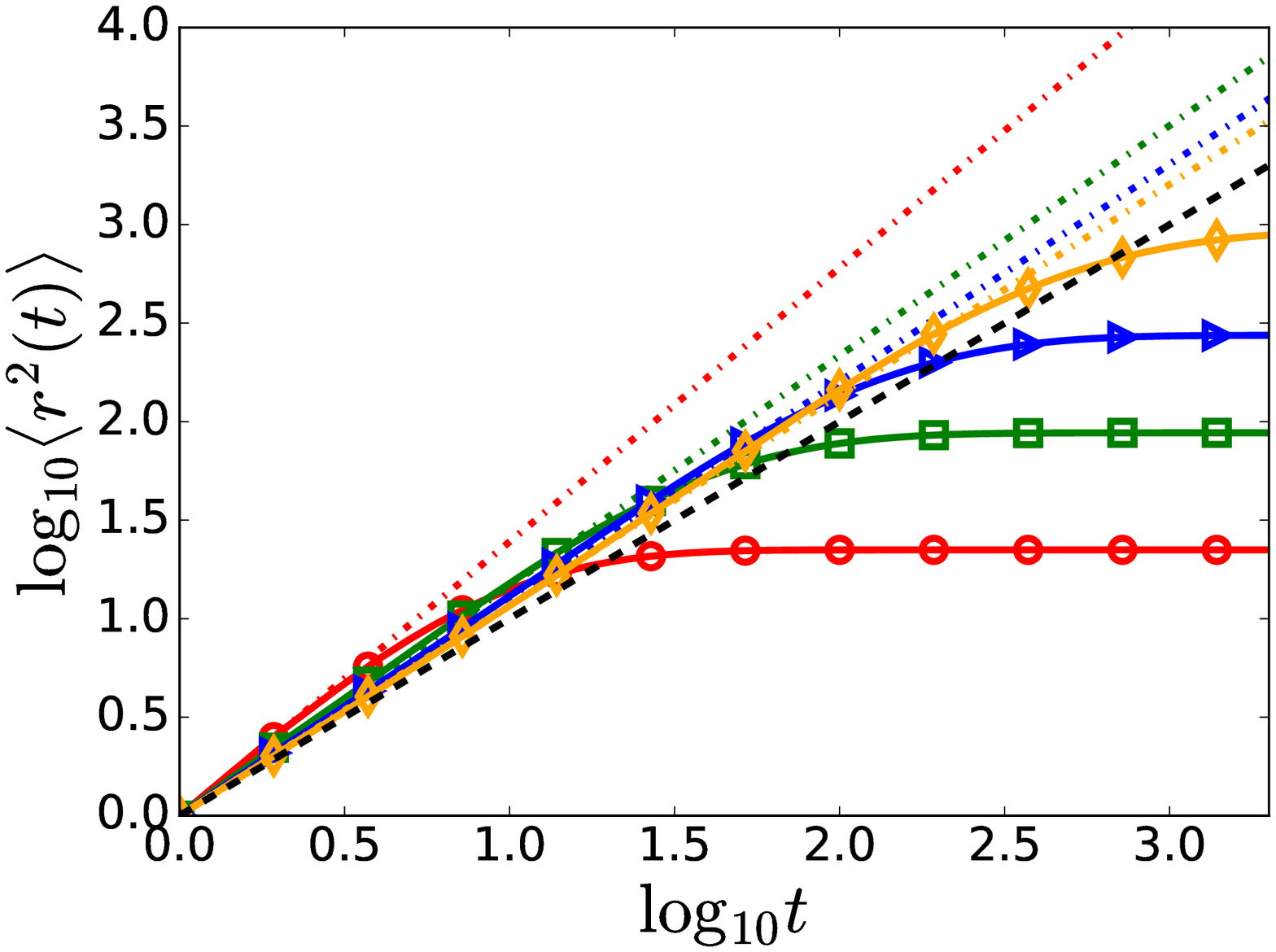}};
        \begin{scope}[x={(image.south east)},y={(image.north west)}]
            \node[anchor=south west,inner sep=0] (image) at (0.18,0.57) {\includegraphics[width=0.175\textwidth]{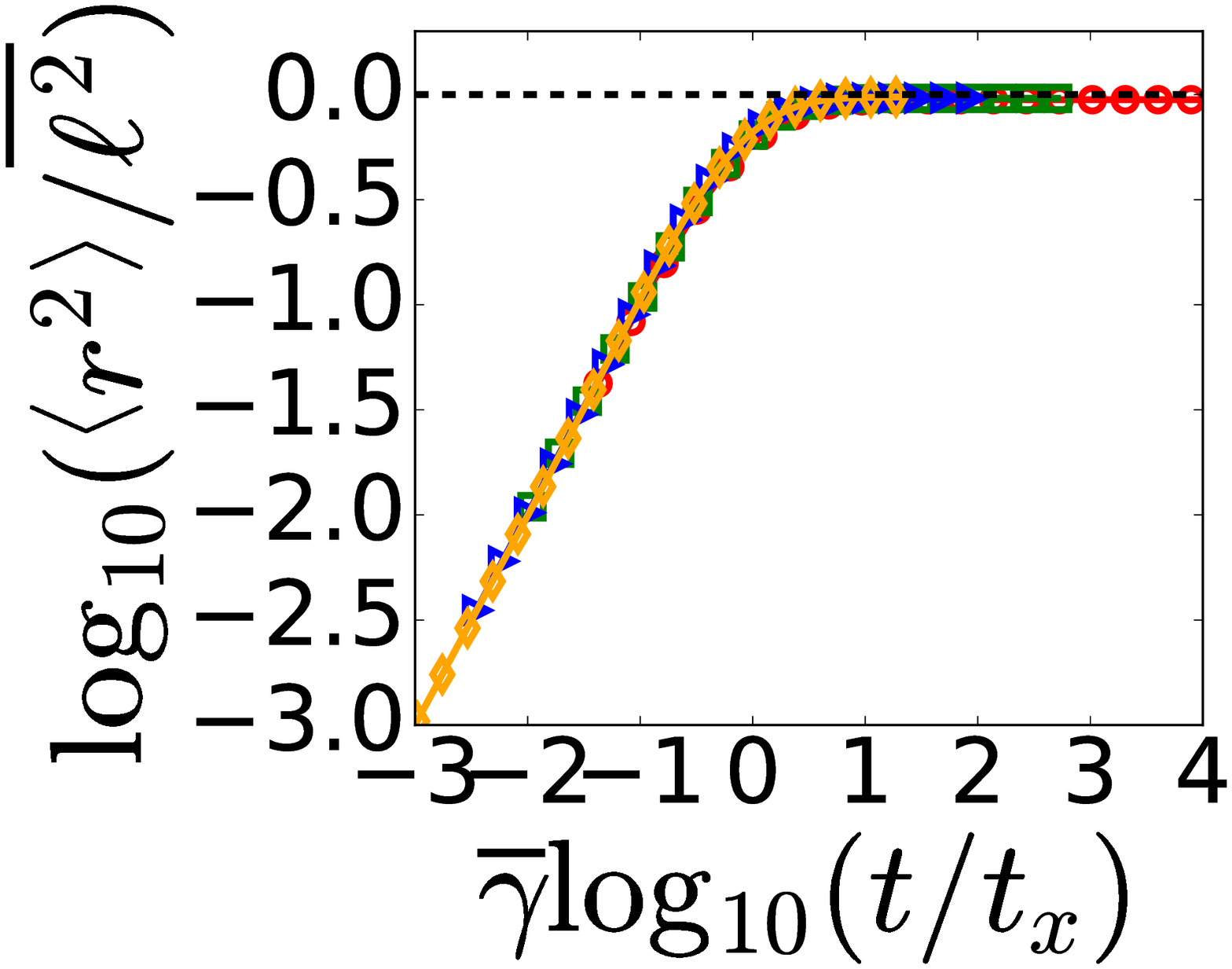}};
        \end{scope}
    \end{tikzpicture}
\label{MSD_variosN_x50}
}
\subfloat[]{
   \begin{tikzpicture}
        \node[anchor=south west,inner sep=0] (image) at (0,0) {\includegraphics[width=0.5\textwidth]{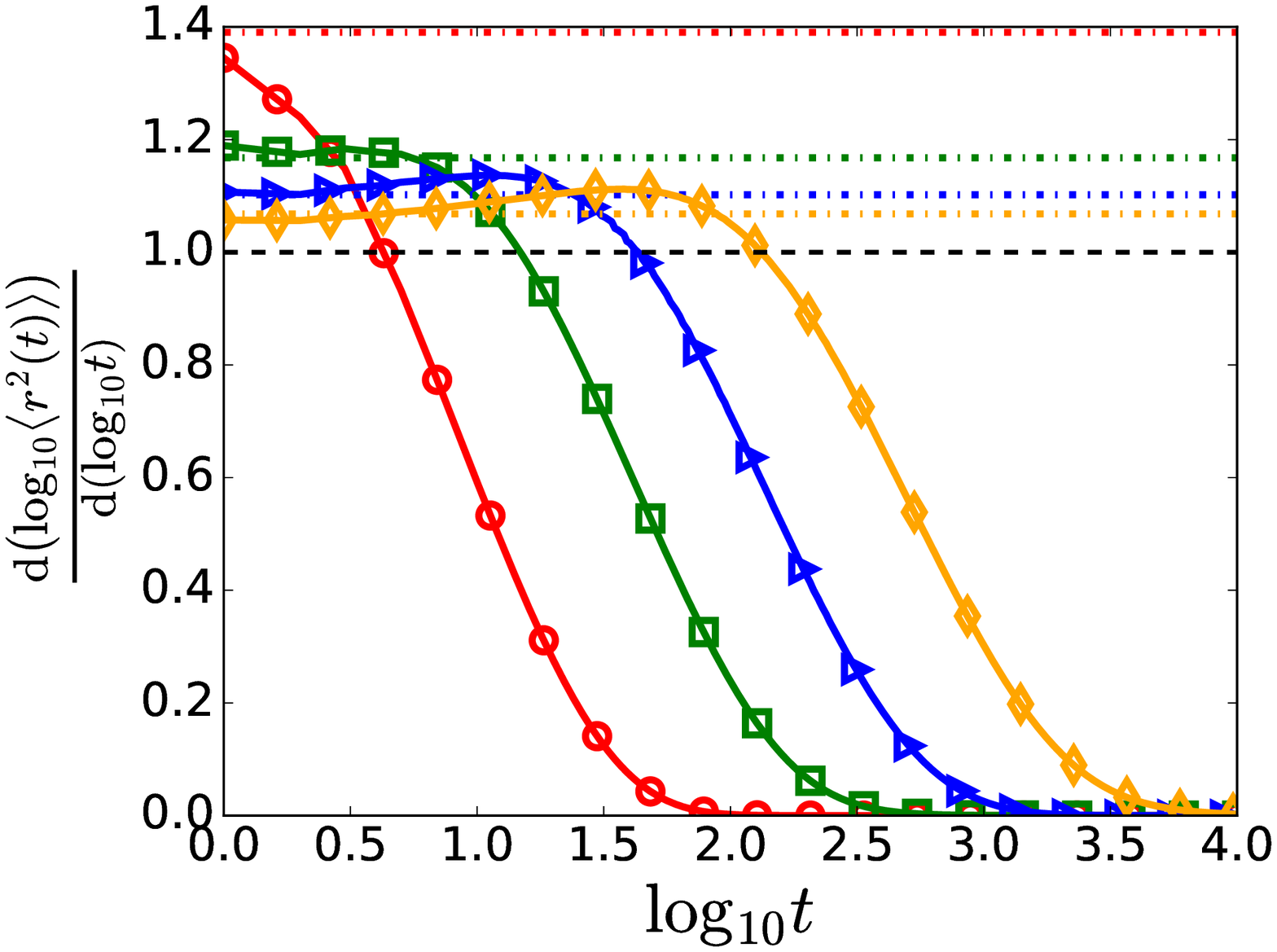}};
        \begin{scope}[x={(image.south east)},y={(image.north west)}]
            \node[anchor=south west,inner sep=0] (image) at (0.20,0.18) {\includegraphics[width=0.18\textwidth]{inset_fig9b}};
        \end{scope}
    \end{tikzpicture}
\label{der_x50_k1_variosN}
}
\caption{(a) Time evolution of $\left \langle r^2(t) \right \rangle$ for NW-networks with $x=50$: $N=100$ (red circles), $N=250$ (green squares), $N=500$ (blue triangles), and $N=1000$ (orange diamonds). The dash-dotted lines indicate results for $t^{\overline{\gamma}}$ [see Eq.~(\ref{scaling_almaas_mod})]. (b) Detail of the numerical derivative of $\log_{10} \left \langle r^2(t) \right \rangle$ with respect to $\log_{10}t$ for the series in Fig.~\ref{MSD_NWs_x50_k1}(a). The insets show the data collapses obtained using the scaling ansatz in Eq.~(\ref{scaling_almaas_mod}). The dash-dotted lines indicate results for ${\overline{\gamma}}$ [see Eq.~(\ref{eq_gamma_approx_eq_1})]. The black dashed line is a guide for the eye to locate a normal diffusion.}
\label{MSD_NWs_x50_k1}
\end{figure}

As before, we begin our discussion by presenting the results for several system sizes, when $k=1$ and $x$ is fixed (see Fig.~\ref{MSD_NWs_x50_k1}). We can see that a clear superdiffusive behavior emerges, before saturation. As expected, for a given $x$, the larger the value of $p$ (i.e., the smaller the system size $N$), the more superdiffusive the system is and the faster it saturates. For instance, in the case of $x=50$, $\gamma_0=1.18$ and $t_{\mathrm{max}} \approx 4$ for $N=250$, whereas  $\gamma_0=1.061$ and $t_{\mathrm{max}} \approx 38$ for $N=1000$. Taking into account these results, as well as those in Subsec.~\ref{subsec_x1} for $k=1$ and in Ref.~\cite{almaas03}, we propose the following ansatz for $k=1$ and $p\ll1$:

\begin{equation}
\left \langle r^2(t) \right \rangle \approx \left\{ \begin{array}{r}
t^{\overline{\gamma}},\\
\overline{\ell^2},
\end{array}\right.\begin{array}{l}
\textnormal{for \ensuremath{t\ll t_x}}\\
\textnormal{for \ensuremath{t\gtrsim t_x}}
\end{array},
\label{scaling_almaas_mod}
\end{equation}

\noindent where $t_x$ denotes the approximated crossover-time to saturation, given by:

\begin{equation}
t_x \approx \left\{ \begin{array}{r}
\overline{\ell^2},\\
 \sqrt{\overline{\ell^2}}\xi,
\end{array}\right.\begin{array}{l}
\textnormal{for \ensuremath{\sqrt{\overline{\ell^2}} \lesssim \xi }}\\
\textnormal{for \ensuremath{\sqrt{\overline{\ell^2}}\gtrsim \xi }}
\end{array},
\label{cross_over_time_eq}
\end{equation}

\noindent and $\overline{\gamma}$ represents an approximated superdiffusion exponent:

\begin{equation}
\overline{\gamma} \approx \gamma_{0}\approx \left\{ \begin{array}{r}
1,\\
\log_{10}\overline{\ell^2}/ \log_{10}t_x,
\end{array}\right.\begin{array}{l}
\textnormal{for \ensuremath{\sqrt{\overline{\ell^2}} \lesssim \xi }}\\
\textnormal{for \ensuremath{\sqrt{\overline{\ell^2}}\gtrsim \xi }}
\end{array}.
\label{eq_gamma_approx_eq_1}
\end{equation}

\noindent Note that Eqs.~(\ref{approx_cycle}) and (\ref{scaling_k1_x1}) can be easily derived from Eqs.~(\ref{scaling_almaas_mod}), (\ref{cross_over_time_eq}) and (\ref{eq_gamma_approx_eq_1}). In the insets of Fig.~\ref{MSD_NWs_x50_k1} we show the excellent collapses we obtained for $0<p\leq0.5$, by using Eq.~(\ref{scaling_almaas_mod}).

\begin{figure}[h!]
\centering
\subfloat[]{
   \begin{tikzpicture}
        \node[anchor=south west,inner sep=0] (image) at (0,0) {\includegraphics[width=0.5\textwidth]{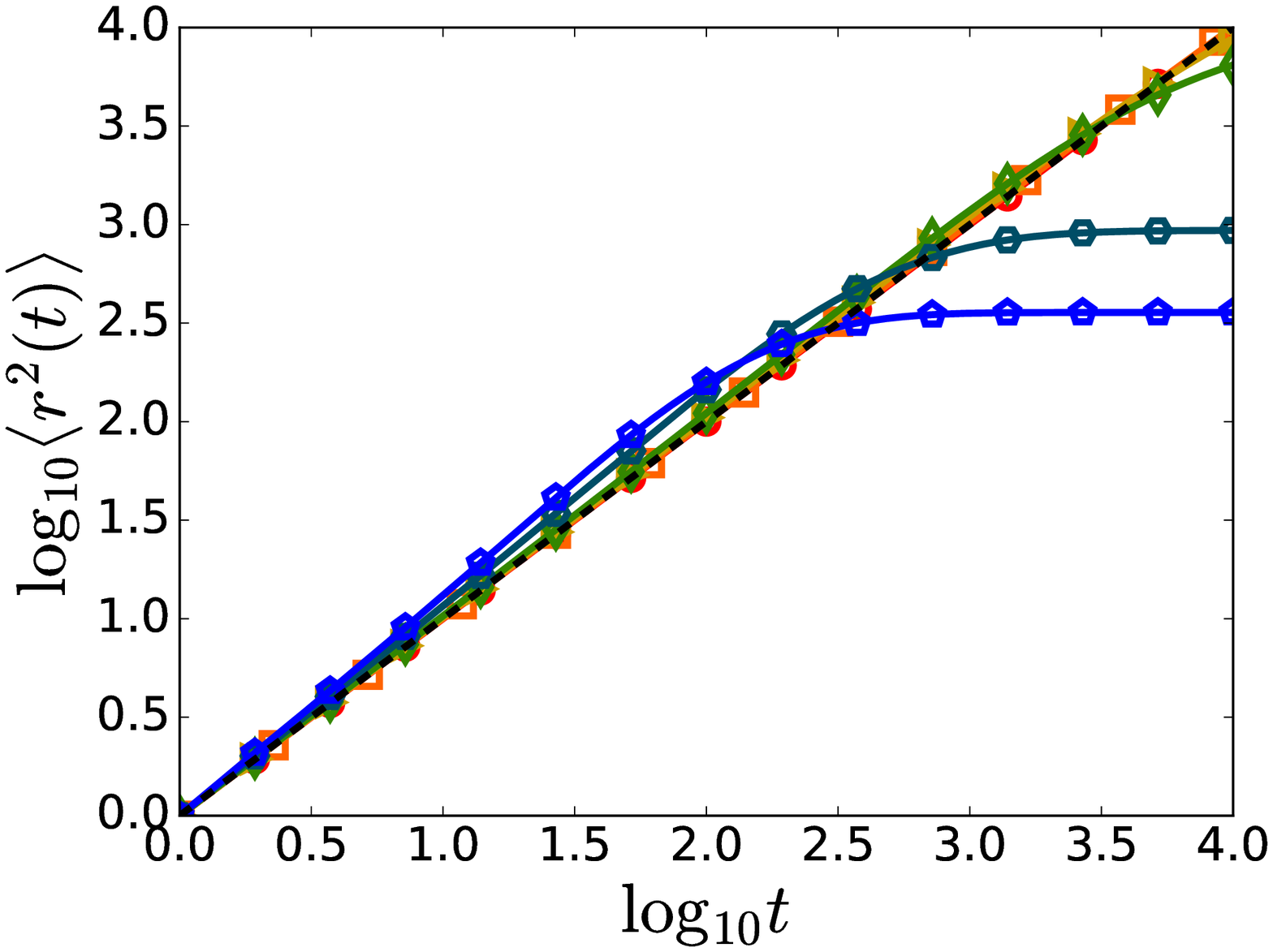}};
        \begin{scope}[x={(image.south east)},y={(image.north west)}]
            \node[anchor=south west,inner sep=0] (image) at (0.57,0.18) {\includegraphics[width=0.18\textwidth]{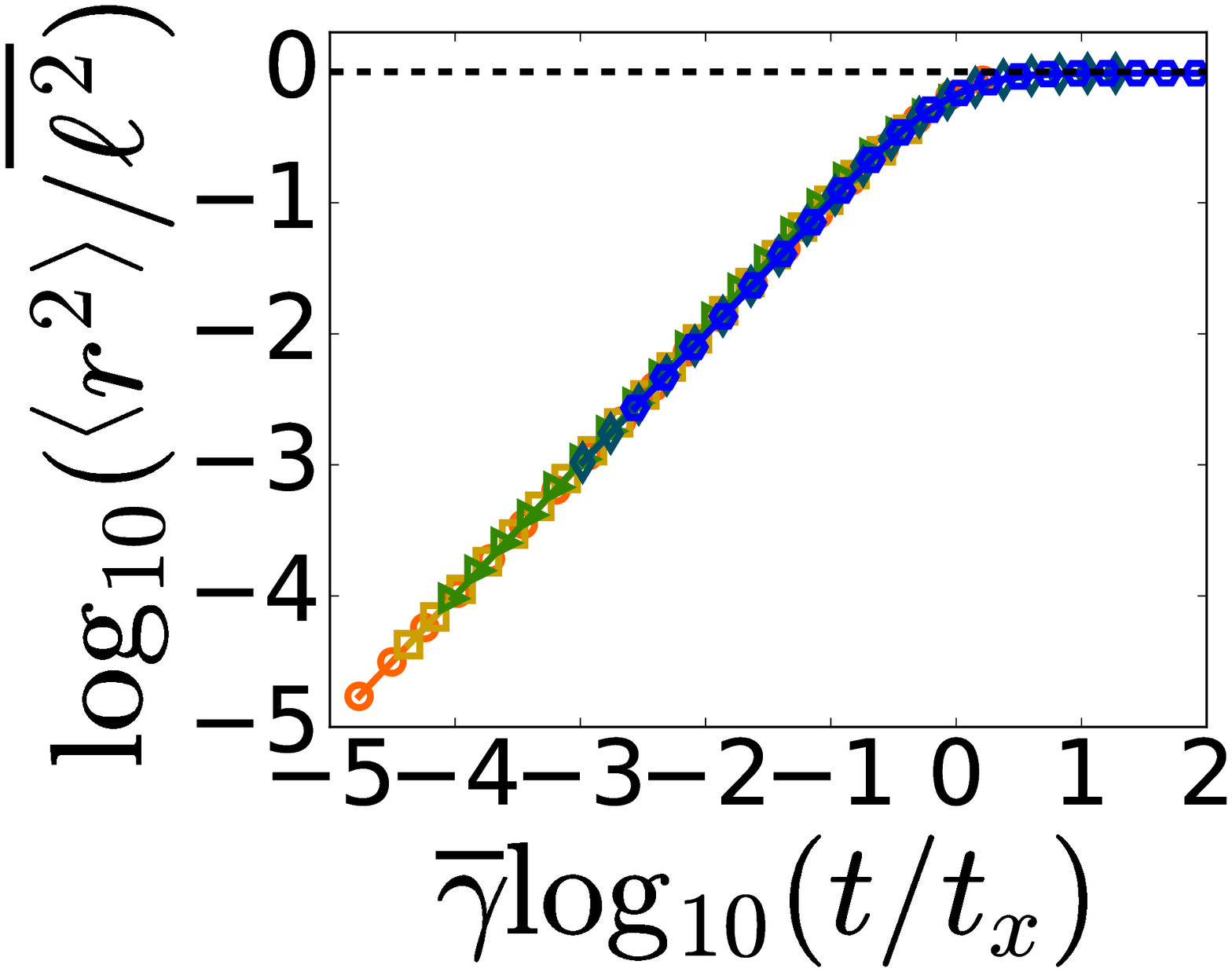}};
        \end{scope}
    \end{tikzpicture}
\label{MSD_variosX_k1}
}
\subfloat[]{
   \begin{tikzpicture}
        \node[anchor=south west,inner sep=0] (image) at (0,0) {\includegraphics[width=0.5\textwidth]{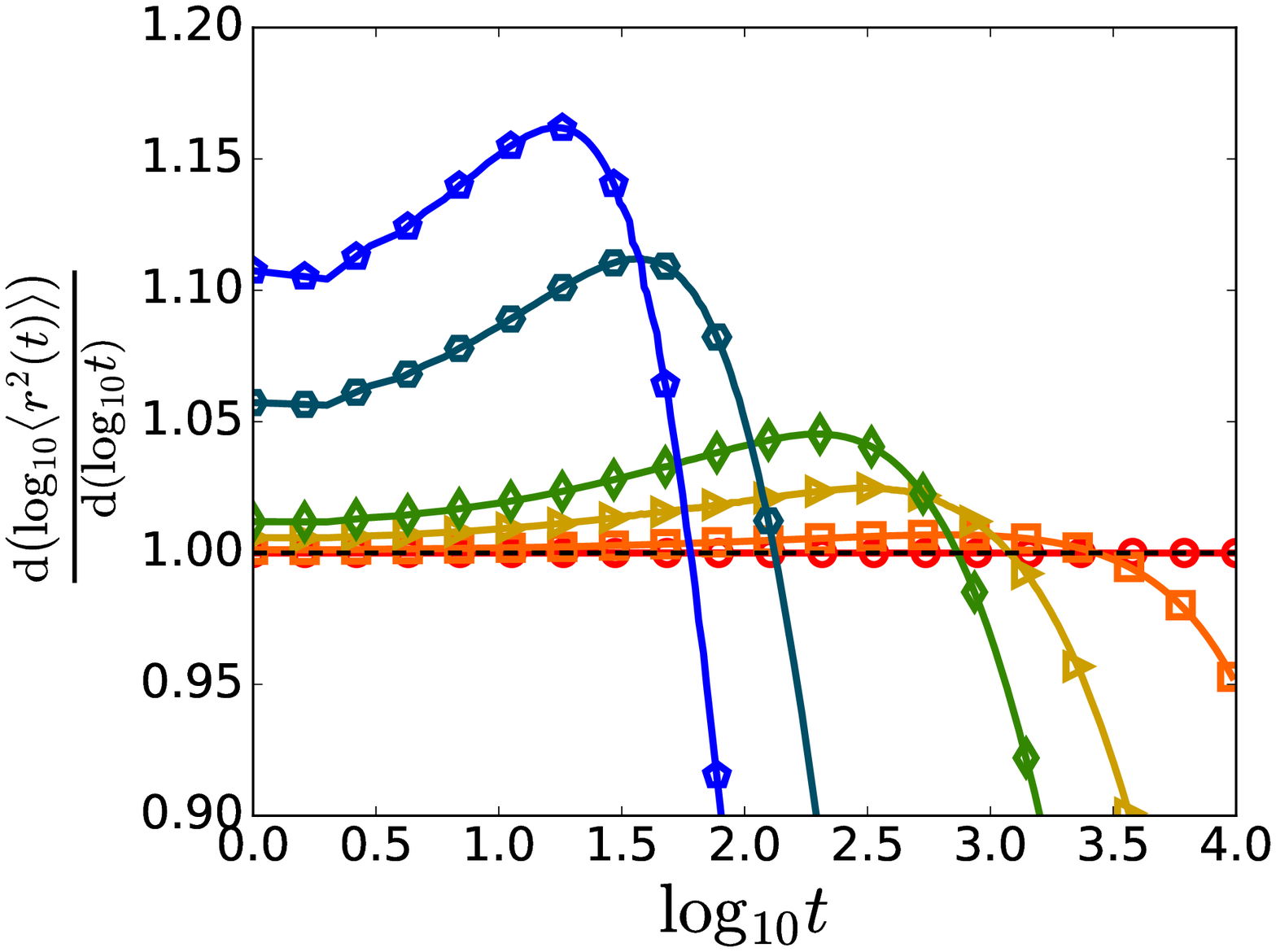}};
        \begin{scope}[x={(image.south east)},y={(image.north west)}]
            \node[anchor=south west,inner sep=0] (image) at (0.59,0.57) {\includegraphics[width=0.175\textwidth]{inset_fig10b}};
        \end{scope}
    \end{tikzpicture}
\label{der_MSD_variosX_k1}
}
\caption{(a) Dependence of the time-evolution of $\left \langle r^2(t) \right \rangle$ for NW-networks with $N=1000$ and $k=1$: $x=0$ (circles), $x=1$ (squares), $x=5$ (triangles), $x=10$ (diamonds), $x=50$ (hexagons) and $x=100$ (pentagons). (b) Detail of the numerical derivative of $\log_{10} \left \langle r^2(t) \right \rangle$ with respect to $\log_{10}t$ for the series in Fig.~\ref{MSD_NW_k1_variosX}(a). The insets shows the collapses obtained by using Eqs.~(\ref{scaling_almaas_mod}), (\ref{cross_over_time_eq}) and (\ref{eq_gamma_approx_eq_1}). The black dashed line is a guide for the eye to locate a normal diffusion.}
\label{MSD_NW_k1_variosX}
\end{figure}

In Fig.~\ref{MSD_NW_k1_variosX}, we show the effect of increasing the number of shortcuts on $\left \langle r^2(t) \right \rangle$ for a given system size $N$, when $k=1$. As expected, the larger the value of $x$, the earlier the walker begins to reach the shortcuts (at average distance $\xi=1/p=N/x$ for $k=1$), and, therefore, the more superdiffusive the arrangement is, before saturation starts. On the other hand, the larger the value of $x$, the smaller the average saturation value of $\left \langle r^2(t) \right \rangle$ ($\overline{\ell^2}$), and, the sooner the finite size effects emerge. It is also worth mentioning that, by using Eq.~(\ref{scaling_almaas_mod}), good data collapse can be obtained when $N$ is fixed and $x$ varies (see insets in Fig.~\ref{MSD_NW_k1_variosX}).

In the case of $0<p\leq 0.01$, the addition of extra-links does not modify the behavior of the Log-derivative described previously for $x=1$. However, as can be seen in Fig.~\subref*{der_MSD_variosX_k1}, when $0.01<p\leq 0.1$, the Log-derivative is not monotonically increasing for $t \leq t_{\mathrm{max}}$: during the first time-steps, it decreases and reaches a minimum. Then it grows until reaching a maximum value associated to $\gamma_{\mathrm{max}}$ at $t=t_{\mathrm{max}}$. However, the mean features of the diffusive behavior remain the same. That is: for a given network size, the larger the amount of extra-links $x$, the larger the values of $\gamma_0$ and $\gamma_{\mathrm{max}}$, and the smaller the duration of the superdiffusive regime, which is proportional to $t_{\mathrm{max}}$.

\begin{figure}[h!]
\centering
\subfloat[]{
\includegraphics[width=0.5\linewidth]{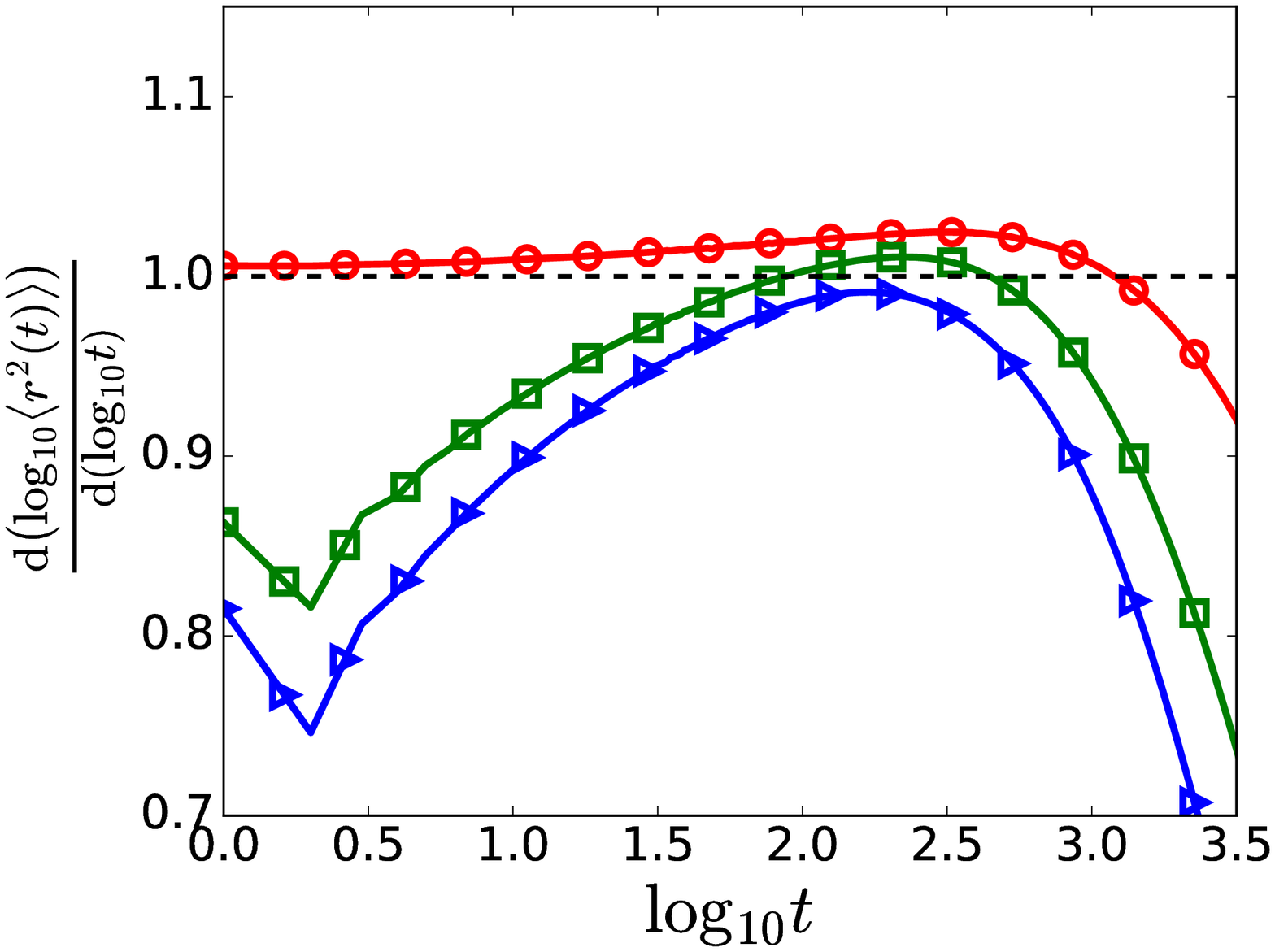}
\label{comp_NW_x5_k1k2}
}
\subfloat[]{
\includegraphics[width=0.5\linewidth]{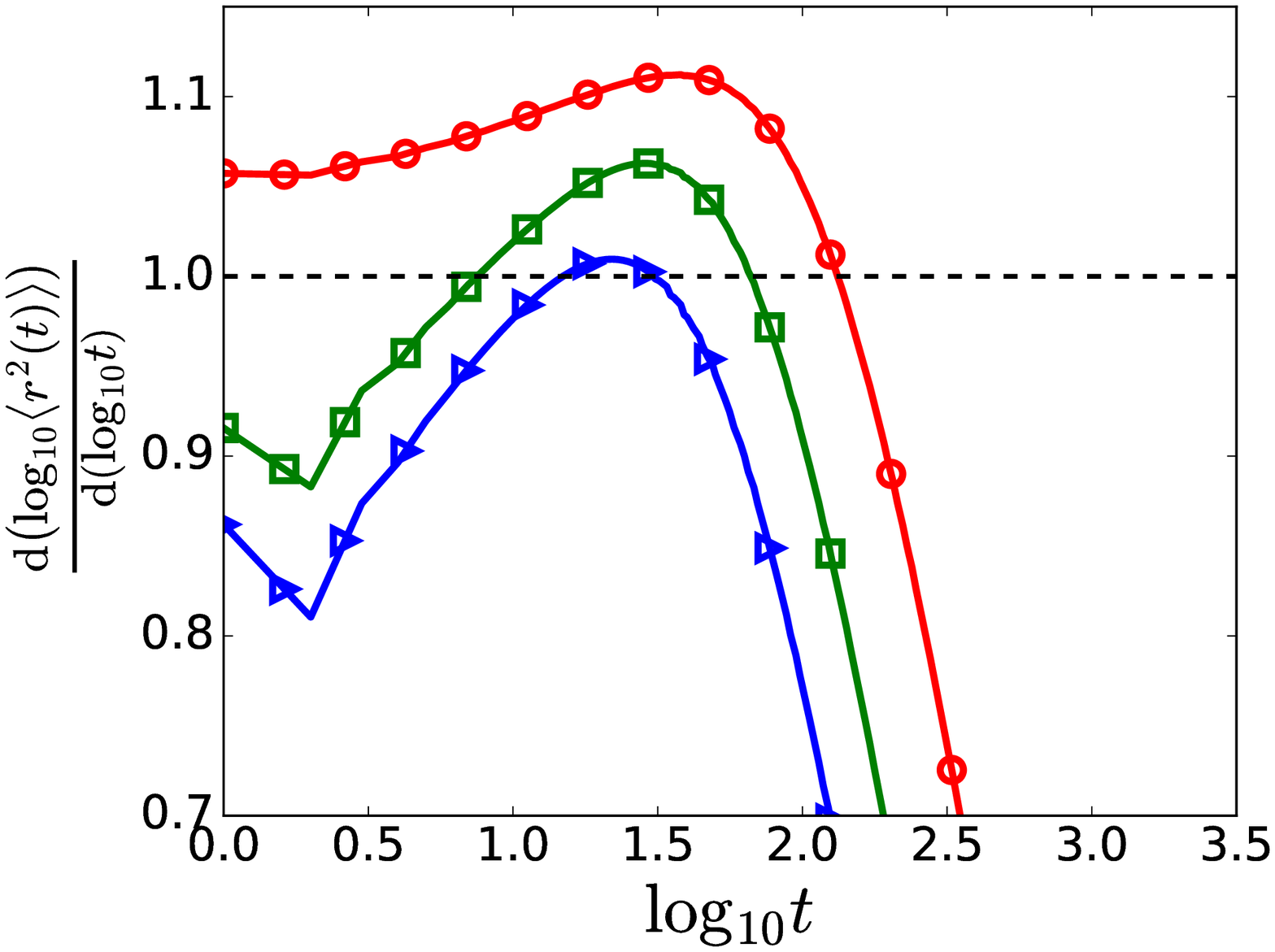}
\label{comp_NW_x50_k1k2}
}
\caption{Detail of the numerical derivative of $\log_{10} \left \langle r^2(t) \right \rangle$ with respect to $\log_{10}t$ for an average NW-network with $N=1000$: $k=1$ (red circles), $k=2$ (green squares) and $k=3$ (blue triangles). (a) $x=5$. (b) $x=50$. The black dashed line is a guide for the eye to locate a normal diffusion.}
\label{comp_NW_x_k1k2}
\end{figure}

Finally, in the case of average NW-networks with $x>1$ and $k>1$, the transitory subdiffusion is present during the first time-steps (see Fig.~\ref{comp_NW_x_k1k2}). As expected, for a given $x$ and $N$, our calculations show that the larger the value of $k$, the more subdiffusive the transitory regime is and the sooner the system saturates. On the other hand, we can see that the addition of extra-links speeds up the diffusion before saturation, no matter the value of $k$. Thus, for a given $N$ and a large enough $x$, superdiffusion may emerge (see  Fig.~\ref{comp_NW_x_k1k2}). However, note that increasing $x$ also makes the finite size effects appear sooner and, therefore, it may hinder superdiffusion materialization.

\subsection{Comparison with previous results for $k=1$}
\label{Compa_Sec}

In this subsection, we compare Eq.~(\ref{scaling_almaas_mod}) with the scaling ansatz in Refs.~\cite{kulkarni00,almaas02,almaas03}:

\begin{equation}
 \left \langle r^2(t) \right \rangle \approx \left\{ \begin{array}{r}
t,\\
\overline{\ell^2},
\end{array}\right.\begin{array}{l}
\textnormal{for \ensuremath{t\ll\xi ^2}}\\
\textnormal{for \ensuremath{t\gtrsim \xi\sqrt{\overline{\ell^2}}}}
\end{array}
\label{scaling_almaas}
\end{equation}

\noindent Equation (\ref{scaling_almaas}) is valid for the sparse regime of NW-networks, with $k=1$ and $p\ll1$, and large size $N$ (such that the terms of $\mathcal{O}(1/N)$ can be omitted). However, the results for $\left \langle r^2(t) \right \rangle$ presented in Ref. \cite{almaas03} refer only to the case $\xi \leq \sqrt{\overline{\ell}}$ and $0<p\leq 0.01$ (i.e., $x \gg 1$). Under those conditions and with the approximation $\overline{\gamma}\approx1$ in Eq.~(\ref{eq_gamma_approx_eq_1}), it is easy to see that Eqs.~(\ref{scaling_almaas_mod}) and (\ref{scaling_almaas}) are identical, and both provide good scaling results. As discussed above, when $0<p\lesssim 0.01$, the exponent $\gamma$ is very close to one, especially in the case of large systems (i.e., $N\gg1000$). On the other hand, if $k=1$ and $p\ll1$, the resulting networks barely match the topological characteristics of the WS-model in Refs.~\cite{watts98,porter12} as their local clustering coefficient is too small (see Fig.~\ref{SW_modelos}).

In the case of $0.01<p\lesssim 0.1$, our findings for small and medium NW-networks show that a clear superdiffusive behavior appears during the first time-steps (see Figs.~\ref{MSD_NWs_x50_k1} and \ref{MSD_NW_k1_variosX}). Indeed, this behavior for larger values of $p$ is similar to the outcome reported in Ref.~\cite{Huang06} for a WS-network with $N=9844$, $p = 0.1$ and average degree $\left \langle s(i) \right \rangle=6$. In these conditions, given that $\gamma>1$, Eq.~(\ref{scaling_almaas}) is no longer accurate, while Eq.~(\ref{scaling_almaas_mod}) provides approximate but good results.

On the other hand, it is worth mentioning that Eq.~(\ref{scaling_almaas}) is not defined for systems in which $\sqrt{\overline{\ell}}\leq \xi$. That is, NW-networks with very few extra-links ($x\sim1$). In contrast, Eq.~(\ref{scaling_almaas_mod}) allows us to obtain an universal curve even for cycle grahs (i.e., when $x=0$ and $\xi=1/p$ diverges).

\begin{figure}[h!]
\centering
\subfloat[]{
   \begin{tikzpicture}
        \node[anchor=south west,inner sep=0] (image) at (0,0) {\includegraphics[width=0.5\textwidth]{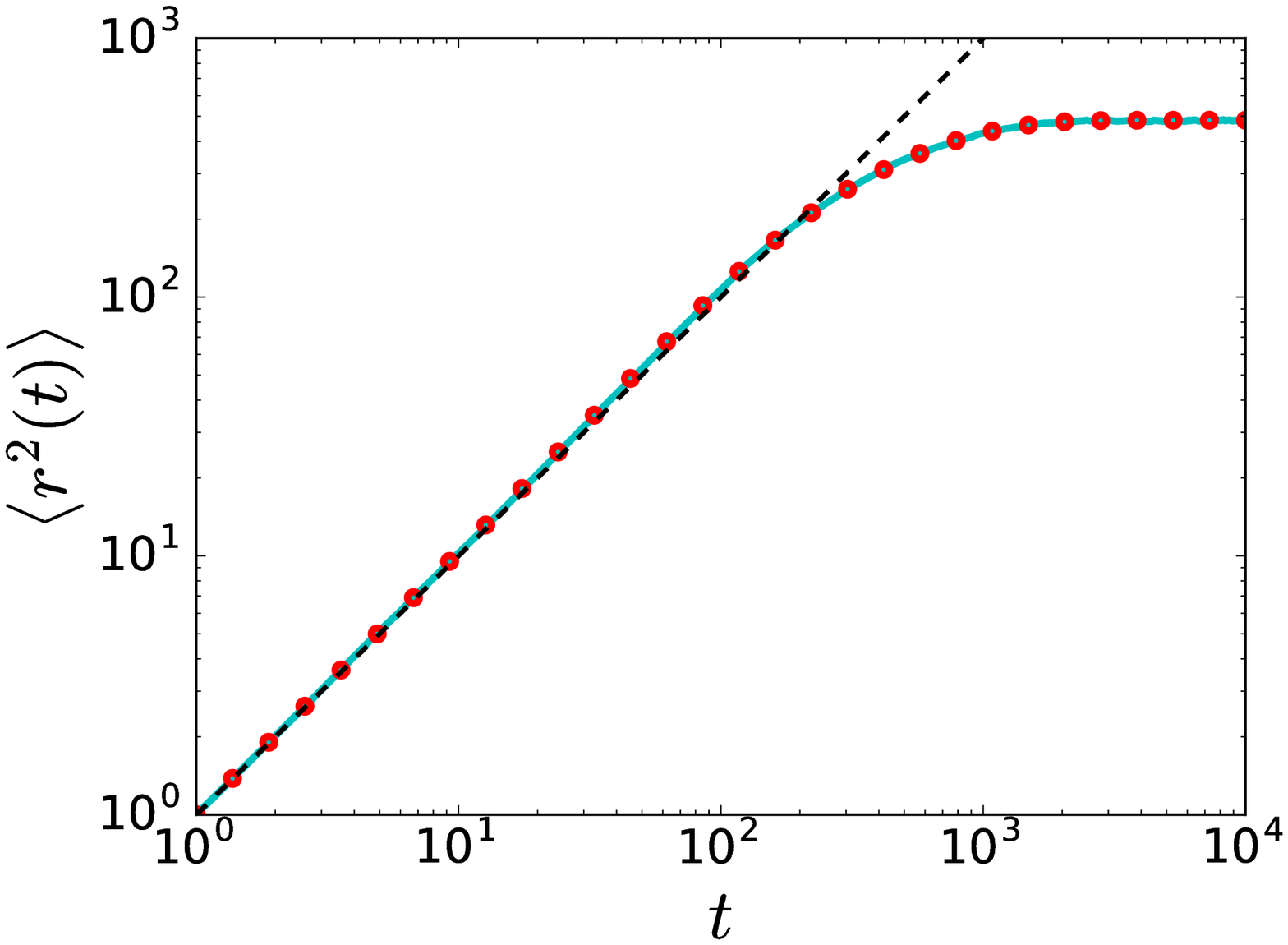}};
        \begin{scope}[x={(image.south east)},y={(image.north west)}]
            \node[anchor=south west,inner sep=0] (image) at (0.52,0.18) {\includegraphics[width=0.21\textwidth]{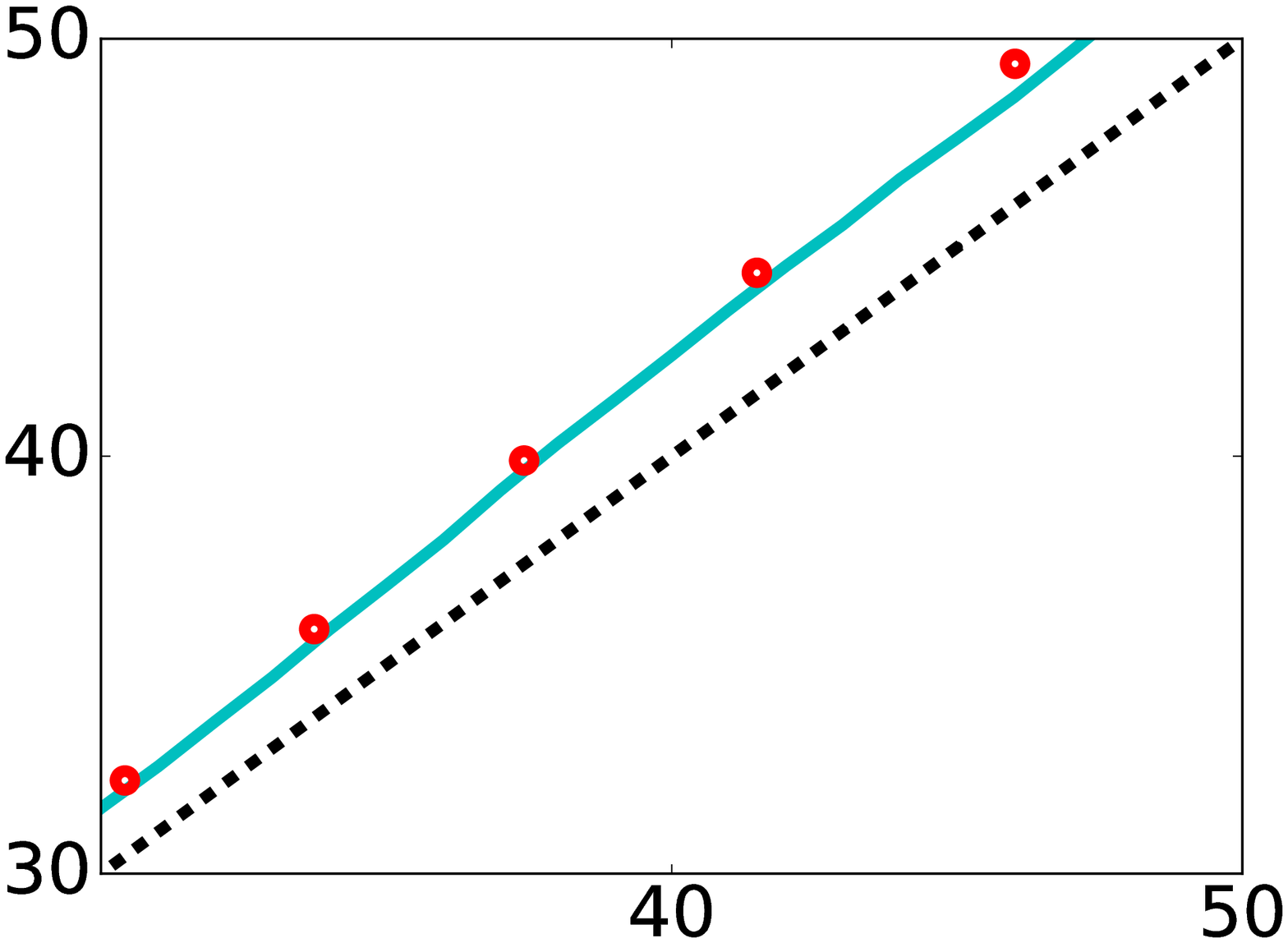}};
        \end{scope}
    \end{tikzpicture}
\label{sim_MSD_EqCG}
}
\subfloat[]{
\centering
\includegraphics[width=0.5\linewidth]{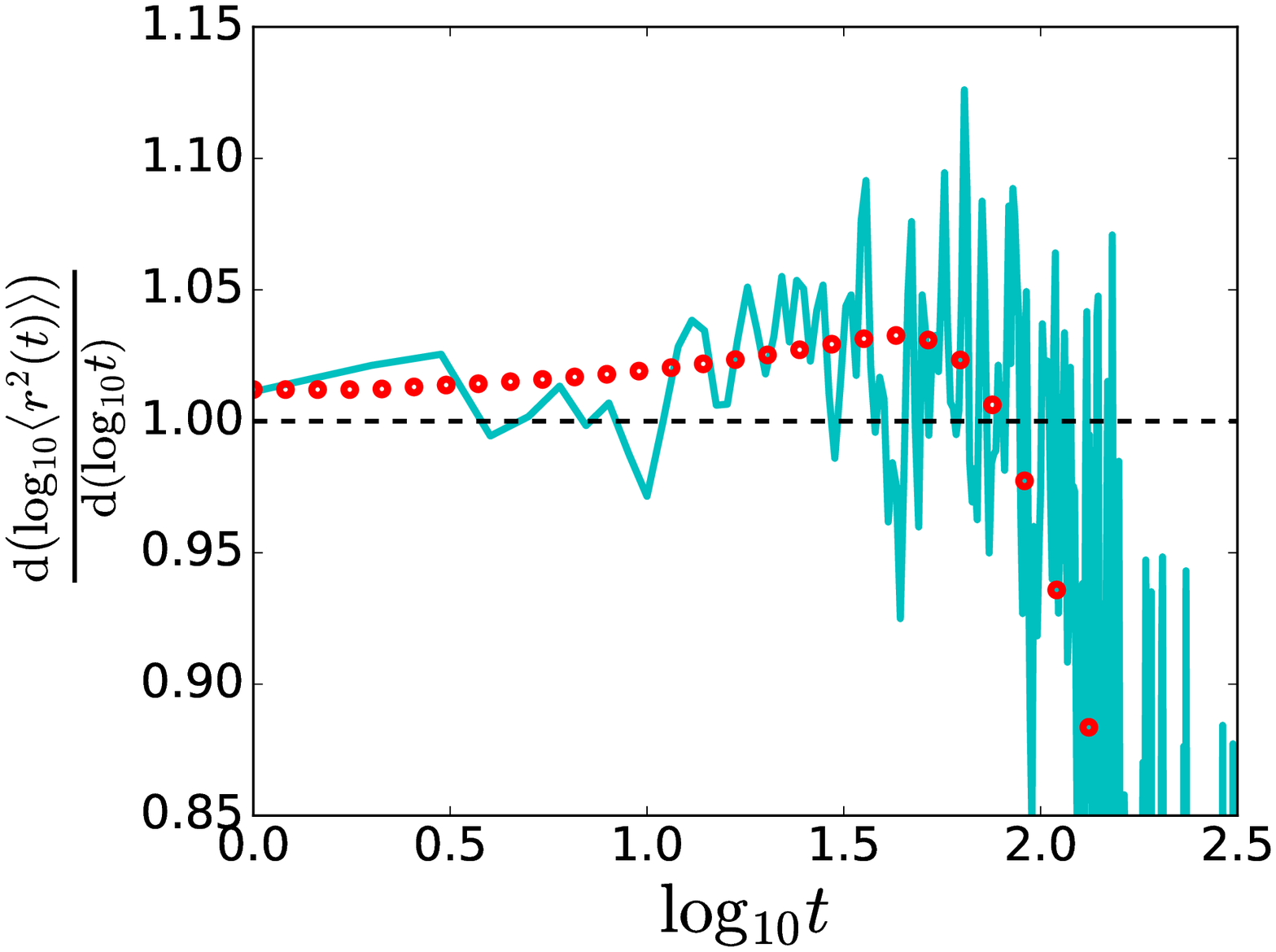}
\label{der_sim_MSD_EqCG}
}
\caption{(a) Time evolution of $\left \langle r^2\left ( t \right ) \right \rangle $ for a $C_{100,1}^+$ with $\Delta=1$ [see Fig.~\ref{redes_offset}(a)]. The red symbols represent the MC results [Eq.~(\ref{eq_MSD_multpl_new})], while the cyan solid line indicates the average result from 50000 independent numeric simulations. The inset shows greater details of the differences between the two series. (b) Detail of the numerical derivative of $\log_{10} \left \langle r^2(t) \right \rangle$ with respect to $\log_{10}t$ of the series shown in Fig.~\ref{sim_MSD_MCGs_100}(a). The black dash-dotted lines in both panels are a guide for the eye to locate a normal diffusion.}
\label{sim_MSD_MCGs_100}
\end{figure}

Finally, it should be noted that the numerical results for $\left \langle r^2(t) \right \rangle$ in Refs.~\cite{almaas02,almaas03} are based in computer simulations not on MC formalism. This difference is significant, since, for any given topology, the characterization provided by the MC framework is numerically exact and, consequently, it is more accurate than the descriptions obtained from computer simulations, even when averaging over a huge amount of realizations (see Fig.~\ref{sim_MSD_MCGs_100} for an example).

\section{Conclusions}
\label{conclusiones}

In this work, we have analyzed the mean-square displacement of a random-walker on NW-networks. Our results were based on the discrete time-evolution of $\left \langle r^2(t) \right \rangle$ obtained by the MC formalism. The characterization provided by this framework is numerically exact. Despite the high computing cost for large system sizes, which forced us to limit the calculations to $N\leq1500$, the highly precise results and scaling analysis clearly indicate that our results should be valid in the very large $N$ limit. To provide a  clear characterization of sub- and supper-diffusive behavior, we systematically obtained the numerical Log-derivatives of $\left \langle r^2(t) \right \rangle$.

In the case of $k=1$ i.e., when each node is initially connected only to its 2 nearest neighbors, our results show that normal diffusion occurs only in the case of cycle graphs, i.e., $x = 0$. Indeed, if $x>0$, the average diffusive behavior of these systems before saturation is always superdiffusive ($\left \langle r^2(t) \right \rangle\sim t^\gamma$ with $\gamma>1$), even during the first time-steps. Indeed, for a given system size $N$, the larger the value of $p$ (i.e., the amount of shortcuts $x$), the more superdiffusive it is, but the sooner it saturates. On the other hand, by analyzing the behavior of $\left \langle r^2(t) \right \rangle$ for all the different NW-networks with $x=1$ and a given size $N$, we have presented evidence that there are topological configurations that exhibit subdiffusion, although on average their behavior is superdiffusive. When the extra-link connects to nodes that are very close, that resulting structure may trap the random walker inside it and hinder its diffusion.

The previous super-diffusive behavior for $k=1$, $x>1$, $N\leq250$, contrasts with the Gaussian diffusion regime
described in Refs.~\cite{almaas02,almaas03} for very large NW-networks with $p\leq0.01$, $x\gg1$ and $\xi \leq \sqrt{\overline{\ell}}$. However, our results also consider that, on increasing $N$, the diffusion exponent $\gamma$ systematically decreases, hinting that for large systems $\left \langle r^2(t) \right \rangle\sim t^\gamma\approx t$ during the first time steps. On the other hand, it is worth mentioning that, according to the generally accepted features of SW in Refs.~\cite{watts98,porter12}, if $k=1$ and $p\ll1$, the resulting arrangements barely match the topological characteristics of the WS-model, as their local clustering is too small.

Taking into account our results for $k=1$, we have proposed a new scaling ansatz for NW-networks in Eqs.~(\ref{scaling_almaas_mod}), (\ref{cross_over_time_eq}) and (\ref{eq_gamma_approx_eq_1}). Using the same conditions assumed in Refs.~\cite{almaas02,almaas03}, it is possible to arrive the scaling ansatz by Almaas \textit{et al.} [Eq.~(\ref{scaling_almaas})]. Additionally, the new scaling can be applied to a wider range of $p$ values, namely: $0.01<p\lesssim 0.1$ and $0\leq p$ (i.e., $\sqrt{\overline{\ell}} \leq \xi$).

On the other hand, when $k>1$, our results for NW-networks indicate that $\left \langle r^2(t) \right \rangle$ exhibits a transitory subdiffusion, during the first time-steps. For a given $x$ and $N$, our calculations show that the larger the value of $k$, the more subdiffusive the transitory regime is and the sooner the system saturates. Indeed, if $N$ is too small, subdiffusion overlaps with saturation. On the other hand, we can see that the addition of extra-links speeds up the diffusion before saturation, no matter the value of $k$. Thus, when $k>1$, superdiffusion may emerge for large enough $x$ and $N$. However, note that increasing $x$ also makes the finite size effects appear sooner and, therefore, it may hinder superdiffusion materialization.

The introduction of this discussion about the influence of the network size and its topology on the mean-square displacement of small-world systems opens new possibilities for characterizing diffusion dynamics in such arrangements. For instance, our findings provide solid ground for discussing the presence of anomalous diffusion in the spreading processes that take place in small communities (such as the spreading of diseases, rumors, or computer virus, among others).

\begin{acknowledgments}

This work was supported by the project MTM2015-63914-P from the Ministry of Economy and Competitiveness of Spain and by the Brazilian agencies CNPq through Grant No. 151466/2018-1 (AA-P) and CAPES. RFSA also acknowledges the support of the National Institute of Science and Technology for Complex Systems (INCT-SC Brazil).
\end{acknowledgments}


\section*{APPENDIX}
\setcounter{subsection}{0}
\label{appendix_A}

In the thermodynamic limit, a random walk on a $(N,k)-$cycle graph is equivalent to the one-dimensional example of a walker performing at each time step $t$ a jump of length $\iota_t=\pm1,\cdots,\pm k$ independently chosen at each time according to a distribution $p(\iota)=1/(2k)$. Its position $j$ after $t$ steps is the sum of $t$ independent displacements:

\begin{equation}
j(t)=\sum_{t=1}^t \iota_t.
\end{equation}

According to $p(\iota)$, its first two moments $\left \langle \iota \right \rangle=0$ and $\left \langle \iota^2 \right \rangle=(1/k)\sum_{i=1}^k i^2=(k+1)(2k+1)/6$ are finite. Thus, following Sec. 1.1 in Ref. \cite{Bouchaud1990}, we obtain Eq.~\ref{Prob_inf_cycle}.

To derive the value of $\left \langle r^2(t) \right \rangle$ for finite $(N,k)-$cycle graphs before saturation (i.e., $10\lesssim t\ll (3N^2)/(2k^2+3k+1)$) in Eq.~\ref{approx_cycle_k2}, we simply readjust the features for one-dimensional lattices to those of $(N,k)-$cycle graphs, namely:
\begin{enumerate}[label=(\roman*)]
\item At time $t$, the \enquote{one-dimensional} position of the walker is $j\in \mathbb{Z}$. In a spatially discrete $(N,k)-$cycle graph, the shortest path distance between the origin of the walk ($j(t=0)=0$) and $j(t)$ is given by $\left \lceil j/k \right \rceil$.
\item The maximum distance between the origin of the walk and the walker is $j_{\max} = \max(j)=kt$ (i.e., $P(j>j_{\max},t)=0$).
\end{enumerate}

\begin{figure}[h!]
\centering
\includegraphics[width=0.5\textwidth]{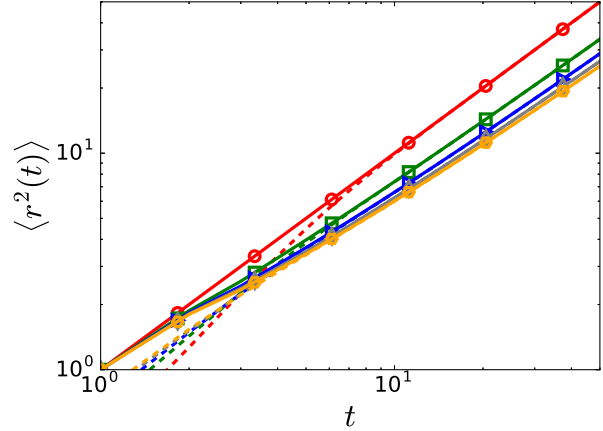}
\caption{Time evolution of $\left \langle r^2(t) \right \rangle$ on $(N,k)-$cycle graphs with distinct $(N,k)$ combinations: $(1000,1)$ (red circles), $(1000,2)$ (green squares), $(1500,3)$ (blue triangles), $(1500,4)$ (grey diamonds), and $(1500,5)$ (orange hexagons). Continuous lines show the results for Eq.~\ref{eq_MSD_multpl_new} and dashed lines represent the results for Eq.~\ref{approx_cycle_k2}.}
\label{cycle_teorico_simulation}
\end{figure}

Finally, in  Fig.~\ref{cycle_teorico_simulation} we compare the time evolution of $\left \langle r^2(t) \right \rangle$ on $(N,k)-$cycle graphs obtained from Eq.~\ref{eq_MSD_multpl_new} with those for Eq.~\ref{approx_cycle_k2}. As can be seen, for large enough $N$ and $10\lesssim t$, they are in excellent agreement.

\end{document}